\documentclass[a4paper,12pt]{article}

\usepackage{cite}

\usepackage{fullpage}
\usepackage[font=footnotesize,labelfont=bf,margin=1cm]{caption}

\usepackage{color}
\usepackage{amsfonts}
\usepackage{amsmath}
\usepackage{amssymb}

\usepackage{multirow}
\usepackage{multicol}

\numberwithin{equation}{section}

\newcommand{\qandq}{\qquad\mathrm{and}\qquad}
\newcommand{\diag}{\mathrm{diag}}
\newcommand{\beq}{\begin{equation}}
\newcommand{\eeq}{\end{equation}}

\newcommand{\dd}{\mathrm{d}}		
\newcommand{\HH}{\mathcal{H}}			
\newcommand{\MM}{\mathcal{M}}			
\newcommand{\LL}{\mathcal{L}}			
\newcommand{\DD}{\mathcal{D}}			
\newcommand{\FF}{\mathcal{F}}
\newcommand{\AAA}{\mathcal{A}}
\newcommand{\g}{\mathfrak{g}}	

\newcommand{\hR}{\hat{R}}
\newcommand{\hg}{\hat{g}}

\newcommand{\hx}{\hat{x}}

\newcommand{\ha}{\hat{a}}
\newcommand{\hta}{\hat{\tilde{a}}}
\newcommand{\hmu}{{\hat{\mu}}}
\newcommand{\hnu}{{\hat{\nu}}}

\newcommand{\bm}{{\bar{m}}}
\newcommand{\bn}{{\bar{n}}}
\newcommand{\bk}{{\bar{k}}}
\newcommand{\bl}{{\bar{l}}}
\newcommand{\bp}{{\bar{p}}}
\newcommand{\bq}{{\bar{q}}}
\newcommand{\btheta}{{\bar{\theta}}}
\newcommand{\bg}{\bar{g}}

\newcommand{\tz}{{\tilde{z}}}
\newcommand{\ta}{\tilde{a}}
\newcommand{\ttheta}{{\tilde{\theta}}}
\newcommand{\ty}{\tilde{y}}
\newcommand{\ttt}{\tilde{t}}
\newcommand{\tu}{\tilde{u}}
\newcommand{\tv}{\tilde{v}}

\newcommand{\dm}{{\dot{m}}}
\newcommand{\dn}{{\dot{n}}}
\newcommand{\dpp}{{\dot{p}}}
\newcommand{\dq}{{\dot{q}}}
\newcommand{\dg}{\dot{g}}

\newcommand{\ua}{{\underline{a}}}
\newcommand{\ub}{{\underline{b}}}

\begin{document}

\begin{titlepage}
\vfill

\begin{flushright}
QMUL-PH-14-19
\end{flushright}

\vfill

\begin{center}
   \baselineskip=16pt
   	{\Large \bf Strings, Branes and the Self-dual Solutions of Exceptional Field Theory}
   	\vskip 2cm
   	{\sc		David S. Berman\footnote{\tt d.s.berman@qmul.ac.uk} and
	 		Felix J. Rudolph\footnote{\tt f.j.rudolph@qmul.ac.uk}}
	\vskip .6cm
    {\small \it Queen Mary University of London, Centre for Research in String Theory, \\
             School of Physics, Mile End Road, London, E1 4NS, England} \\ 
	\vskip 2cm
\end{center}

\begin{abstract}
It has been shown that membranes and fivebranes are wave-like or monopole-like solutions in some higher dimensional theory. Here the picture is completed by combining the wave and monopole solutions into a single solution of Exceptional Field Theory. This solution solves the twisted self-duality constraint. The 1/2 BPS brane spectrum, consisting of fundamental, solitonic and Dirichlet branes and their bound states, in ten- and eleven-dimensional supergravity may all be extracted from this single solution of Exceptional Field Theory. The solution's properties such as its asymptotic behaviour at the core and at infinity are investigated. 

\end{abstract}

\vfill

\setcounter{footnote}{0}
\end{titlepage}

\tableofcontents

\section{Introduction}
Duality symmetries have been at the heart of developments in string theory. A duality --- the presence of a hidden symmetry or relation between theories --- once found, immediately provokes a set of questions. The very presence of the duality seems to  imply a lack of understanding of the theory; one hopes to discover the reason for the hidden symmetry and perhaps discover a theory in which this duality symmetry is manifest. 

Certainly since 1995, this idea has been very successful in the context of supersymmetric field theories. The S-duality present in $\mathcal{N}=4$ super Yang-Mills theory and in the low energy effective description of the $\mathcal{N}=2$ super Yang-Mills theory is explained by the realization that these theories can be described as coming from the dimensional reduction of a single theory, with $(0,2)$ supersymmetry in six dimensions (in M-theory terms, the theory of the M-theory fivebrane)\cite{Verlinde:1995mz,Witten:1995zh,Witten:1997sc}. This has led to a profound exploration of field theories that have used the $(0,2)$ theory to explain all manner of field theory properties in lower dimensions and has even gone as far as providing an explanation of the esoteric duality of the Geometric Langlands program \cite{Kapustin:2006pk}.

Central to these explanations is the fact that the six-dimensional theory is equipped with a self-dual three-form field strength. The source for this field comes from a self-dual string. The coupling in this theory is of order one as it must be for a self-dual theory. Thus there is no notion of perturbation theory. To generate a perturbative regime we must introduce a scale into the theory so that at given energies we can form a small dimensionless parameter with which we can do perturbative calculations. Most simply this is achieved by compactifying the six dimensional theory on a circle of radius $R$. This then produces five-dimensional Yang-Mills with coupling given by $R$. 

The more interesting thing happens when we compactify on a torus and the reduced theory is four-dimensional Yang-Mills with coupling given by the complex structure of the torus. The S-duality in the four-dimensional theory is then just a consequence of the modular invariance of the torus, i.e. a trivial consequence of the geometric description of the torus. The self-dual strings of the six dimensional theory can wrap either the $a$ or $b$ cycles of the torus. This describes the 1/2 BPS spectrum of the $\mathcal{N}=4$ theory that in turn forms a representation of the $SL(2)$ duality group \cite{Witten:1995zh}. The $SL(2)$ duality symmetry is then just from relabelling the cycles on the torus that the string wraps. The string itself is self-dual and has no duality property; the duality is emergent based on different perspectives in the reduction of the theory. This idea has been used and studied in numerous applications and directions. In what follows, we will describe something like a supergravity analogue.

Exceptional Field Theory \cite{Hohm:2013pua, Hohm:2013vpa, Hohm:2013uia, Hohm:2014fxa} was developed as a theory to make manifest the U-duality groups of M-theory. The theory lives in a space with many new dimensions and comes equipped with something known as the \emph{physical section condition} which determines how one may carry out a reduction of the theory back down to eleven dimensions or less. The U-duality groups enter when due to the presence of isometries in the extended space there are different ways to do this reduction and usual spacetime as embedded in this theory becomes ambiguous. Thus the duality is a consequence of an ambiguity in the description of the reduction. Now what of 1/2 BPS states in the theory? Brane and string solutions of supergravity all transform under U-duality. We will describe a single object in Exceptional Field Theory and show how the ambiguity in its reduction leads to all the 1/2 BPS objects in supergravity that transform into each other under U-duality. 

This object obeys a twisted self-duality constraint in terms of the gauge fields of Exceptional Field Theory\footnote{The twisted self-duality equation was first described in the seminal work of Cremmer, Julia and Pope \cite{Cremmer:1998px}.}. It will also be geometrically self-dual in the following sense. The solution is heuristically speaking a superposition of a wave and monopole with a single free quantized parameter which gives both the monopole charge and the wave momentum. We call this self-dual because if one reduces this object so that the wave propagates along some circle then this will give rise to an electric charge from the reduced perspective which is equal to the magnetic charge coming from the monopole. Thus in some sense it is a lift of a self-dual KK-dyon\cite{Khuri:1995xq}. This object is the analogue of the self-dual string in the $(0,2)$ theory. Its reduction provides us with the complete 1/2 BPS spectrum and the action of the duality group on the BPS spectrum is just a relabelling in terms of this single self-dual object.

We will analyse this solution and make explicit its reduction to the various branes in supergravity (such as fundamental strings, solitonic branes and D-branes). We also show how the EFT solution (which maybe thought of as a superposition of a wave and monopole) changes its character as one moves from asymptotic infinity towards the core of the solution. 

In this paper we will begin with an EFT primer that should give a sufficient outline to follow the results presented here. For the original works see \cite{Hohm:2013pua, Hohm:2013vpa, Hohm:2013uia, Hohm:2014fxa}. Then we describe the solution and subsequently its reduction to the various branes in supergravity. We analyse the behaviour of the self-dual solution in EFT and in particular look at the difference between the core and asymptotic regions and conclude with a discussion on the absence of singularities of the EFT solution.

For the relevant literature the reader may consult recent reviews of DFT and related theories given in \cite{Aldazabal:2013sca, Berman:2013eva,Hohm:2013bwa}; the primary work of Siegel in constructing a duality manifest formalism \cite{Siegel93a,Siegel93b,Siegel93c}; the work outlining DFT \cite{Hull:2009mi, Hull:2009zb, Hohm:2010jy, Hohm:2010pp}. The steps towards an M-theory equivalent to DFT for the truncated theory are given in \cite{Hull:2007zu, Pacheco:2008ps, Hillmann:2009ci, Berman:2010is, Coimbra:2011ky, Coimbra:2012af, Berman:2011pe, Berman:2011kg, Berman:2011cg, Berman:2011jh, Berman:2012vc, Park:2013gaj, Cederwall:2013oaa, Cederwall:2013naa, Strickland-Constable:2013xta, Park:2014una}. And there is now a whole host of interesting works in this field, for a representative but by no means complete sample one may start with \cite{Lee:2014mla, Hull:2014mxa, Coimbra:2014qaa, Rosabal:2014rga, Hohm:2014xsa, Hohm:2014sxa, Cederwall:2014kxa, Cederwall:2014opa, Blair:2014zba}. All of this is related to the long standing $E_{11}$ program of West and collaborators, see for example \cite{West:2001as, Englert:2003zs, West:2003fc, Kleinschmidt:2003jf, West:2004kb}.

\section{Exceptional Field Theory}

The primary idea behind Exceptional Field Theory \cite{Hohm:2013pua} is to make the exceptional symmetries of eleven-dimensional supergravity manifest. The appearance of the exceptional groups in dimensionally reduced supergravity theories was first discussed in \cite{Cremmer:1997ct, Cremmer:1998px}. In Exceptional Field Theory one first performs a decomposition of eleven-dimensional supergravity but with no reduction or truncation into an $(11-D) \times D$ split. That is one takes the eleven dimensions of supergravity to be
\begin{equation}
M^{11} = M^{11-D} \times M^D \, .
\end{equation}
Then one supplements the $D$ so called ``internal'' directions with additional coordinates to linearly realize the exceptional symmetries. This follows the previous works on truncated theories that realize the exceptional duality groups \cite{Hull:2007zu, Pacheco:2008ps, Hillmann:2009ci, Berman:2010is, Coimbra:2011ky, Coimbra:2012af} through introducing novel extra dimensions and leads to
\begin{equation}
M^{11} \longrightarrow M^{11-D} \times M^{\dim E_D}
\end{equation}
where $\dim E_D$ is the dimension of the fundamental\footnote{This is true for $D=6,7,8$. For other groups a different representation might be needed for the construction, e.g. for $D=4$ where $E_4=SL(5)$ it is the {\bf{10}}.} representation of the exceptional group $E_D$ and $M^{\dim E_D}$ is a coset manifold that comes equipped with the coset metric of ${E_D}/{H}$ (where H is the maximally compact subgroup of $E_D$). This ``exceptional extended geometry'' has been constructed for several U-duality groups but was previously restricted to truncations of the eleven-dimensional theory where the ``external'' metric was taken to be flat and off-diagonal terms (the ``gravi-photon'') were set to zero. Furthermore, coordinate dependence was restricted to the internal extended coordinates.

Exceptional Field Theory provides the full, non-truncated theory which allows for a dependence on all coordinates, external, internal and extended. This allows for eleven-dimensional supergravity to be embedded into a theory that is fully covariant under the exceptional groups $E_D$ for $D=6,7,8$ \cite{Hohm:2013vpa,Hohm:2013uia,Hohm:2014fxa} (a supersymmetric extension for $D=6,7$ can be found in \cite{Musaev:2014lna,Godazgar:2014nqa}) and more recently also for $D=4,5$ \cite{Blair:2014zba, Abzalov:2015ega}. 

It is worthwhile at this stage to describe how the U-duality groups become related to the embedding of the eleven dimensions in the extended space. The combination of p-form gauge transformations and diffeomorphism give rise to a continuous local $E_D$ symmetry. This however is not U-duality which is a global discrete symmetry that only occurs in the presence of isometries. (See \cite{Berman:2014jba} for the equivalent discussion for DFT). Crucially however there is also a {\it{physical section condition}} that provides a constraint in EFT that restricts the coordinate dependence of the fields to a subset of the dimensions and thus there naturally appears a physical submanifold which we identify as usual spacetime. When there are no isometries present this section condition constraint produces a canonical choice of how spacetime is embedded in the extended space.  However, in the presence of isometries there is an ambiguity in how one identifies the submanifold in the extended space. This ambiguity is essentially the origin of U-duality with different choices of spacetime associated to U-duality related descriptions. (This is discussed in detail for the case of DFT in \cite{Berman:2014jsa}).

In this paper we focus on the $E_7$ group and the corresponding EFT. We will give a brief overview of the most important concepts of the theory, closely following \cite{Hohm:2013uia} where all the details can be found. We choose $E_7$ since it has all the complexities that we wish to explore. It is expected that the narrative of this paper could easily be repeated for other choices of duality group.

\subsection{Basics of the $E_7$ EFT}
The $E_7$ Exceptional Field Theory lives in a $4+56$-dimensional spacetime. The four dimensional external space has coordinates $x^\mu$ and metric $g_{\mu\nu}={e_\mu}^\ua{e_\nu}^\ub\eta_{\ua\ub}$ which may be expressed in terms of a vierbein. The 56-dimensional extended internal space has coordinates $Y^M$ which are in the fundamental representation of $E_7$. This exceptional extended space is equipped with a generalized metric $\MM_{MN}$ which parametrizes the coset $E_7/SU(8)$.

From earlier work \cite{Hull:2007zu, Pacheco:2008ps, Hillmann:2009ci, Coimbra:2011ky, Coimbra:2012af, Berman:2011jh} the 56-dimensional exceptional extended geometry is known. The generalized tangent space is isomorphic to the sum 
\begin{equation}
TM \oplus \Lambda^2 T^*M \oplus \Lambda^5 T^*M \oplus (T^*M\otimes \Lambda^7 T^*M) 
\end{equation}
where $M$ is the seven-space. The terms in the sum correspond to brane charges: momentum, membrane, fivebrane and KK-monopole charge. These are the conjugate variables to the 56 coordinates $Y^M$ which can be seen as brane \emph{wrapping} coordinates in analogy to the string \emph{winding} coordinates of DFT.

In addition to the external metric $g_{\mu\nu}$ and the generalized metric $\MM_{MN}$, EFT also requires a generalized gauge connection ${\AAA_\mu}^M$ and a pair of two-forms $B_{\mu\nu\ \alpha}$ and $B_{\mu\nu\ M}$ to describe all degrees of freedom of eleven-dimensional supergravity. Here $\alpha=1,\dots,144$ labels the adjoint and $M=1,\dots,56$ the fundamental representation of $E_7$. For more on the nature of these two-forms see \cite{Hohm:2013uia}. For the main part of this paper they will both be zero and not play a role in what follows though they are of course crucial for the consistency of the theory.

Thus, the field content of the $E_7$ exceptional field theory is
\begin{equation}
\left\{ g_{\mu\nu}, \MM_{MN}, {\AAA_\mu}^M, B_{\mu\nu\ \alpha}, B_{\mu\nu\ M}\right\} \, .
\end{equation}
All these fields are then subjected to the physical section condition which picks a subspace of the exceptional extended space. This section condition can be formulated in terms of the $E_7$ generators $(t_\alpha)^{MN}$ and the invariant symplectic form $\Omega_{MN}$ of $Sp(56)\supset E_7$ as
\begin{align}
(t_\alpha)^{MN}\partial_M\partial_N \Phi &= 0 \, , &
(t_\alpha)^{MN}\partial_M\Phi\partial_N \Psi &= 0 \, , &
\Omega^{MN}\partial_M\Phi\partial_N \Psi &= 0
\label{eq:section}
\end{align}
where $\Phi,\Psi$ stand for any field and gauge parameter. 

The equations of motion describing the dynamics of the fields can be derived from the following action
\begin{equation}
\begin{aligned}
S = \int \dd^4x \dd^{56}Y e&\left[\hR 
		+ \frac{1}{48}g^{\mu\nu}\DD_\mu\MM^{MN}\DD_\nu\MM_{MN} \right. \\
		&\quad \left.	- \frac{1}{8}\MM_{MN}\FF^{\mu\nu\ M}{\FF_{\mu\nu}}^N
		- V(\MM_{MN},g_{\mu\nu}) + e^{-1}\LL_{\mathrm{top}}\right] \, .
\end{aligned}
\label{eq:action}
\end{equation}
The first term is a covariantized Einstein-Hilbert term given in terms of the spin connection $\omega$ of the vierbein ${e_\mu}^\ua$ (with determinant $e$)
\begin{equation}
\LL_{\mathrm{EH}}=e\hR =e {e_\ua}^\mu{e_\ub}^\nu\hR_{\mu\nu}{}^{\ua\ub} \qquad \mathrm{where} \qquad
\hR_{\mu\nu}{}^{\ua\ub} \equiv R_{\mu\nu}{}^{\ua\ub}[\omega] + {\FF_{\mu\nu}}^M e^{\ua\rho}\partial_M{e_\rho}^\ub \, .
\end{equation}
The second term is a kinetic term for the generalized metric $\MM_{MN}$ which takes the form of a non-linear gauged sigma model with target space $E_7/SU(8)$. The third term is a Yang-Mills-type kinetic term for the gauge vectors ${\AAA_\mu}^M$ which are used to define the covariant derivatives $\DD_\mu$. The fourth term is the ``potential'' $V$ built from internal extended derivatives $\partial_M$
\begin{equation}
\begin{aligned}
V = &-\frac{1}{48}\MM^{MN}\partial_M\MM^{KL}\partial_N\MM_{KL}
		+ \frac{1}{2}\MM^{MN}\partial_M\MM^{KL}\partial_L\MM_{NK} \\
	& - \frac{1}{2}\g^{-1}\partial_M\g\partial_N\MM^{MN} 
		- \frac{1}{4}\MM^{MN}\g^{-1}\partial_M\g\g^{-1}\partial_N\g
		- \frac{1}{4}\MM^{MN}\partial_M g^{\mu\nu}\partial_N g_{\mu\nu}
\end{aligned}
\end{equation}
where $\g = e^2 = \det g_{\mu\nu}$. The last term is a topological Chern-Simons-like term which is required for consistency. 

All fields in the action depend on all the external and extended internal coordinates. The derivatives $\partial_M$ appear in the non-abelian gauge structure of the covariant derivative and together with the two-forms $B_{\mu\nu}$ in the field strengths ${\FF_{\mu\nu}}^M$.

The gauge connection ${\AAA_\mu}^M$ allows for the theory to be formulated in a manifestly invariant way under generalized Lie derivatives. The covariant derivative for a vector of weight $\lambda$ is given by
\begin{equation}
\begin{aligned}
\DD_\mu V^M &= \partial_\mu V^M - {\AAA_\mu}^K\partial_K V^M
	+ V^K\partial_K {\AAA_\mu}^M 
	+ \frac{1-2\lambda}{2} \partial_K{\AAA_\mu}^K V^M \\
	&\quad + \frac{1}{2}\left[24(t^\alpha)^{MN}(t^\alpha)_{KL}+\Omega^{MN}\Omega_{KL}\right]
		\partial_N{\AAA_\mu}^K V^L \, .
\end{aligned}
\end{equation}
The associated non-abelian field strength of the gauge connection, defined as
\begin{equation}
\begin{aligned}
{F_{\mu\nu}}^M &\equiv 2\partial_{[\mu}{\AAA_{\nu]}}^M 
	- 2{\AAA_{[\mu}}^N\partial_N{\AAA_{\nu]}}^M \\
	&\quad - \frac{1}{2}\left[24(t^\alpha)^{MN}(t^\alpha)_{KL}-\Omega^{MN}\Omega_{KL}\right]
		{\AAA_{[\mu}}^K\partial_N{\AAA_{\nu]}}^L \, ,
\end{aligned}
\end{equation}
is not covariant with respect to vector gauge transformations. In order to form a properly covariant object we extend the field strength with St\"uckelberg-type couplings to the compensating two-forms $B_{\mu\nu\ \alpha}$ and $B_{\mu\nu\ M}$ as follows
\begin{equation}
{\FF_{\mu\nu}}^M = {F_{\mu\nu}}^M - 12(t^\alpha)^{MN}\partial_N B_{\mu\nu\ \alpha} 
		- \frac{1}{2}\Omega^{MN}B_{\mu\nu\ N}  \, .
\label{eq:genfieldstrength}
\end{equation}
For a detailed derivation and explanation of this we refer to \cite{Hohm:2013uia}. The Bianchi identity for this generalized field strength is
\begin{equation}
3\DD_{[\mu}{\FF_{\nu\rho]}}^M = -12(t^\alpha)^{MN}\partial_N\HH_{\mu\nu\rho\ \alpha}
	- \frac{1}{2}\Omega^{MN}\HH_{\mu\nu\rho\ N}
\end{equation}
which also defines the three-form field strengths $\HH_{\mu\nu\rho\ \alpha}$ and $\HH_{\mu\nu\rho\ M}$. The final ingredient of the theory are the twisted self-duality equations for the 56 EFT gauge vectors ${\AAA_\mu}^M$
\begin{equation}
{\FF_{\mu\nu}}^M = \frac{1}{2}e \epsilon_{\mu\nu\rho\sigma} \Omega^{MN} \MM_{NK} \FF^{\rho\sigma\ K} 
\label{eq:selfduality}
\end{equation}
which relate the 28 ``electric'' vectors to the 28 ``magnetic'' ones. This self-duality relation is a crucial property of the $E_7$ EFT and is essential for the results presented here. In fact this sort of twisted self-duality equation has been described many years ago in the seminal work of \cite{Cremmer:1998px}. 

To conclude this brief overview of exceptional field theory, we note that the bosonic gauge symmetries uniquely determine the theory. They are given by the generalized diffeomorphisms of the external and extended internal coordinates. For more on the novel features of the generalized diffeomorphisms in exceptional field theory see \cite{Hohm:2013uia}.

An immediate simplification to the above equations presents itself when the coordinate dependence of fields and gauge parameters is restricted. In Section \ref{sec:solution} we will consider a solution of EFT which only depends on external coordinates. Thus any derivative of the internal extended coordinates, $\partial_M$, vanishes trivially. Furthermore, our solution comes with zero two-form fields $B_{\mu\nu\ \alpha}$ and $B_{\mu\nu\ M}$, thus simplifying the gauge structure further. The upshot of this is a drastic simplification of the theory: covariant derivatives $\DD_\mu$ reduce to ordinary partials $\partial_\mu$, the generalized field strength ${\FF_{\mu\nu}}^M$ is simply given by $2\partial_{[\mu}{\AAA_{\nu]}}^M$, the covariantized Einstein-Hilbert term reduces to the ordinary one and the potential $V$ of the generalized metric vanishes. Finally the Bianchi identity reduces to the usual $\dd \FF^M =0$.  

We will comment in future work on how to reinstate a dependence on internal extended coordinates and thus localize solutions in the exceptional extended space \cite{bermanrudolphlocalised}.

\subsection{Embedding Supergravity into EFT}
\label{sec:embedding}
Having outlined the main features of the $E_7$ EFT, we proceed by showing how eleven-dimensional supergravity can be embedded in it (again following \cite{Hohm:2013uia} closely). Applying a specific solution of the section condition \eqref{eq:section} to the EFT produces the dynamics of supergravity with its fields rearranged according to a $4+7$ Kaluza-Klein coordinate split.

The appropriate solution to the section condition is related to a decomposition of the fundamental representation of $E_7$ under its maximal subgroup $GL(7)$
\begin{equation}
\mathbf{56} \rightarrow 7 + 21 + 7 + 21
\end{equation}  
which translates to the following splitting of the extended internal coordinates
\begin{equation}
Y^M = (y^m, y_{mn}, y_m, y^{mn})
\label{eq:gencoordM}
\end{equation}
where $m=1,\dots,7$ and the pair $mn$ is antisymmetric. We thus have indeed $7+21+7+21=56$ coordinates. The section condition is solved by restricting the coordinate dependence of fields and gauge parameters to the $y^m$ coordinates. We thus have
\begin{equation}
\begin{aligned}
\partial^{mn} &\rightarrow 0 \, , & 
\partial^m &\rightarrow 0 \, , & 
\partial_{mn} &\rightarrow 0 \\
{B_{\mu\nu}}^{mn} &\rightarrow 0 \, , & 
{B_{\mu\nu}}^m &\rightarrow 0 \, , & 
{B_{\mu\nu}}_{mn} &\rightarrow 0 
\end{aligned}
\end{equation}
where the second line is the necessary consequence for the compensating two-form ${B_{\mu\nu}}^M$.

The complete procedure to embed supergravity into EFT can be found in \cite{Hohm:2013vpa,Hohm:2013uia}, here we will focus on those aspects relevant to our results. The Kaluza-Klein decomposition of the eleven-dimensional spacetime metric takes the following form
\begin{equation}
\hg_{\hmu\hnu} = 
			\begin{pmatrix}
				\hg_{\mu\nu} & \hg_{\mu n} \\
				\hg_{m\nu} & \hg_{mn}
			\end{pmatrix} = 
			\begin{pmatrix}
				g_{\mu\nu} + {A_\mu}^m{A_\nu}^ng_{mn} & {A_\mu}^mg_{mn} \\
				g_{mn}{A_\nu}^n & g_{mn}
			\end{pmatrix}
\end{equation}
where hatted quantities and indices are eleven-dimensional. The four-dimensional external sector with its metric $g_{\mu\nu}$ is carried over to the EFT. The seven-dimensional internal sector is extended to the 56-dimensional exceptional space and the internal metric $g_{mn}$ becomes a building block of the generalized metric $\MM_{MN}$. The KK-vector ${A_\mu}^m$ becomes the $y^m$-component of the EFT vector ${\AAA_\mu}^M$. 

The gauge potentials $C_3$ and $C_6$ of supergravity are also decomposed under the $4+7$ coordinate split. Starting with the three-form, there is the purely external three-form part $C_{\mu\nu\rho}$ which lives in the external sector. The purely internal scalar part $C_{mnp}$ is included in $\MM_{MN}$. The one-form part $C_{\mu\ mn}$ is the $y_{mn}$-component of ${\AAA_\mu}^M$. The remaining two-form part $C_{\mu\nu\ m}$ gets encoded in compensating two-form $B_{\mu\nu\ M}$. Similarly for the six-form, the purely internal scalar $C_{m_1\dots m_6}$ is part of $\MM_{MN}$. The one-form $C_{\mu\ m_1\dots m_5}$ is dualized on the internal space and forms the $y^{mn}$-component of ${\AAA_\mu}^M$. The remaining components of $C_6$ with a mixed index structure (some of which need to be dualized properly) are encoded in the two-forms $B_{\mu\nu\ \alpha}$ and $B_{\mu\nu\ M}$.

In the next section we will work with supergravity solutions where the gauge potentials only have a single non-zero component which will be of the one-form type under the above coordinate split. There will not be any internal scalar parts or other mixed index components. The above embedding of supergravity fields into EFT can therefore be simply summarized as follows. The spacetime metric $g_{\mu\nu}$ of the external sector is carried over; the generalized metric $\MM_{MN}$ of the extended internal sector is given in terms of the internal metric $g_{mn}$ by
\begin{equation}
\MM_{MN}(g_{mn}) = g^{1/2}\diag [ g_{mn}, g^{mn,kl}, g^{-1}g^{mn}, g^{-1}g_{mn,kl}]
\label{eq:genmetricSUGRA}
\end{equation}
where the determinant of the internal metric is denoted by $g=\det g_{mn}$, the four-index objects are defined by $g_{mn,kl}=g_{m[k}g_{l]n}$ and similarly for the inverse; and the components of the EFT vector potential ${\AAA_\mu}^M$ are
\begin{align}
{\AAA_\mu}^m 	&= {A_\mu}^m \, , &
\AAA_{\mu\ mn}	&= C_{\mu\ mn}  \, , &
{\AAA_\mu}^{mn}	&= \frac{1}{5!}\epsilon^{mn\ m_1\dots m_5} C_{\mu\ m_1\dots m_5} \, .
\label{eq:vecpotSUGRA}
\end{align}
The final component, $\AAA_{\mu\ m}$, is related to the dual graviton and has no appearance in the supergravity picture, see \cite{Hohm:2013uia}.

It is also possible to embed the Type II theories in ten dimensions into EFT. The Type IIA embedding follows from the above solution to the section condition by a simple reduction on a circle. In contrast, the Type IIB embedding requires a different, inequivalent solution to the section condition \cite{Hohm:2013uia}. Both Type II embeddings are presented in Appendix \ref{app:embedding}

We are now equipped with the tools to relate Exceptional Field Theory to eleven-dimensional supergravity and the Type IIA and Type IIB theory in ten dimensions. This will be useful when analysing the EFT solution we are presenting next.

\section{A Self-dual Solution in EFT}
\label{sec:solution}

Having introduced Exceptional Field Theory with its field content and equations, we can now consider  specific field configurations which solve these equations. We are looking for a solution from which the known supergravity solutions can be extracted by a suitable choice of section. Furthermore, as argued in the introduction, this solution needs to satisfy the twisted self-duality equation \eqref{eq:selfduality}.

Now consider the following set of fields. We take the external sector to be four-dimensional spacetime with one timelike direction $t$ and three spacelike directions $w^i$ with $i=1,2,3$. The external metric is that of a point-like object, given in terms of a harmonic function of the transverse coordinates by
\begin{equation}
g_{\mu\nu} = \diag [-H^{-1/2},H^{1/2}\delta_{ij}]\, , \qquad H(r) = 1 + \frac{h}{r}
\label{eq:exmetric}
\end{equation}
where $r^2=\delta_{ij}w^iw^j$ and $h$ is some constant (which will be interpreted later). 

The 56-dimensional extended internal sector uses the coordinates $Y^M$ given in \eqref{eq:gencoordM}. The EFT vector potential ${\AAA_\mu}^M$ of our solution has ``electric'' and ``magnetic'' components (from the four-dimensional spacetime perspective) that are given respectively by
\begin{equation}
{\AAA_t}^M 	= \frac{H-1}{H} a^M \qquad\mathrm{and}\qquad
{\AAA_i}^M	= A_i \ta^M \, ,
\label{eq:vecpot}
\end{equation}
where $A_i$ is a potential of the magnetic field. The magnetic potential obeys a BPS-like condition where its curl is given by the gradient of the harmonic function that appears in the metric
\begin{equation}
\vec{\nabla}\times\vec{A} = \vec{\nabla} H 
\qquad\mathrm{or}\qquad
\partial_{[i}A_{j]} = \frac{1}{2}{\epsilon_{ij}}^k\partial_k H \, .
\label{eq:AH}
\end{equation}
The index $M$ in ${\AAA_\mu}^M$ labels the 56 vectors, only two of which are non-zero for our solution. The vector $a^M$ in the extended space (a scalar form a spacetime point of view) points in one of the 56 extended directions. Later we will interpret this direction as the direction of propagation of a wave or momentum mode. The dual vector $\ta^M$ denotes the direction dual to $a^M$ given approximately by $a^M \sim \Omega^{MN}\MM_{NK}\ta^K$. This sense of duality between directions of the extended space will be formalized in Section \ref{sec:selfduality}.

Using the relation between $H$ and $A_i$, one can immediately check that ${\AAA_\mu}^M$ satisfies the twisted self-duality equation \eqref{eq:selfduality}. Loosely speaking, the duality on the external spacetime via $\epsilon_{\mu\nu\rho\sigma}$ exchanges electric ${\AAA_t}^M$ and magnetic ${\AAA_i}^M$ components of the potential. The symplectic form $\Omega_{MN}$ acts on the extended internal space and swaps $a^M$ with its dual $\ta^M$. If one goes through the calculation carefully, one sees that minus signs and factors of powers of $H$ only work out if both actions on the external and extended internal sector are carried out simultaneously. We will show this explicitly in Section \ref{sec:selfduality}.

The generalized metric of the extended internal sector, $\MM_{MN}$, is a diagonal matrix with just four different entries, $\{ H^{3/2},H^{1/2},H^{-1/2},H^{-3/2}\}$. The first and last one appear once each, the other two appear 27 times each. The precise order of the 56 entries of course depends on a coordinate choice, but once this is fixed it characterizes the solution together with the choice of direction for $a^M$.

For definiteness, let's fix the coordinate system and pick a direction for $a^M$ which we call $z$, i.e. $a^M=\delta^{Mz}$. The dual direction is denoted by $\tz$ and we have $\ta^M=\delta^{M\tz}$. Then $\MM_{zz}=H^{3/2}$ and $\MM_{\tz\tz}=H^{-3/2}$. For completeness, the full expression for the generalized metric for the coordinates in \eqref{eq:gencoordM} is\footnote{Here $\delta_n$ denotes an $n$-dimensional Kronecker delta.}
\begin{equation}
\MM_{MN} = \diag[ H^{3/2}, H^{1/2}\delta_6, H^{-1/2}\delta_6, H^{1/2}\delta_{15},
				H^{-3/2}, H^{-1/2}\delta_6, H^{1/2}\delta_6, H^{-1/2}\delta_{15} ] \, .
\label{eq:genmetric}
\end{equation}
The second 28 components are the inverse of the first 28 components, reflecting the split of the EFT vector ${\AAA_\mu}^M$ into 28 ``electric'' and 28 ``magnetic'' components.

To get the fields for any other direction, one simply has to perform a $SO(56)$ rotation in the extended space. The rotation matrix $R \in SO(56)$ rotates $a^M$ in the desired direction ${a'}^M$ and at the same time transforms $\MM_{MN}$ according to
\begin{align}
{a'}^M &= {R^M}_N a^N \, , & 
\MM'_{MN} &= {R_M}^K\MM_{KL}{R^L}_N \, .
\label{eq:rot}
\end{align}
Since the action and the self-duality equation is invariant under such a transformation, the fields can freely be rotated in the extended space.

The remaining fields of the theory, namely the two-form gauge fields $B_{\mu\nu\ \alpha}$ and $B_{\mu\nu\ M}$, are trivial. Also the external part of the three-form potential, $C_{\mu\nu\rho}$, vanishes for our solution. This will eventually restrict somewhat the possible supergravity solutions obtained from this EFT solution. Dropping these restrictions would be interesting and provide a technical challenge to repeat this paper but include other fluxes such as those on the external space.

To recap, the fields $g_{\mu\nu}$, ${\AAA_\mu}^M$ and $\MM_{MN}$ as given in equations \eqref{eq:exmetric}, \eqref{eq:vecpot} and \eqref{eq:genmetric} (together with \eqref{eq:AH}) form our solution to EFT. They satisfy the self-duality equation and their respective equations of motion.

Note that all fields directly or indirectly depend on the harmonic function $H$ which in turn only depends on the external transverse coordinates $w^i$. There is thus no coordinate dependence on any of the internal or extended coordinates.  This solution therefore is de-localized and smeared over all the internal extended directions. It is an interesting open question to look at solutions localized in the extended space. In theory, EFT can handle coordinate dependencies on \emph{all} coordinates, even the extended ones. We leave this for future work.

\subsection{Interpreting the Solution}

How do we interpret this solution in Exceptional Field Theory? Before we do this let us return to how solutions in the truncated theory may be interpreted. A wave whose momentum is in a {\it{winding}} direction describes a brane associated with that winding direction, e.g. a wave with momentum along $y_{12}$ describes a membrane extended over the $y^1,y^2$ directions. A monopole-like solution --- by which we mean a Hopf fibration --- where the $S^1$ fibre is a winding direction describes the S-dual brane to that winding direction, e.g. if the fibre of the monopole is $y_{12}$ then the solution describes a fivebrane. Thus in the extended (but truncated) theory branes can have either a description as monopole or as a wave. These statements were the conclusions of \cite{Berkeley:2014nza,Berman:2014jsa}. 

Now because of the truncation it was not possible to describe a given solution in both ways within the same description of spacetime. The key point of EFT is that there is no truncation and so such things are possible. The self-duality relation is simply the Kaluza-Klein description of a solution that has both momentum and non-trivial Hopf fibration, i.e. it is simultaneously electric and magnetic from the point of view of the KK-graviphoton. As commented in the introduction, these are not just solutions to some linear abelian theory but full solutions to the gravitational theory (or in fact EFT). As such they are exact self-dual solutions to the non-linear theory though are charged with respect to some $U(1)$ symmetry that is given by the existence of the $S^1$ in extended space. Our intuition should be shaped by this experience with Kaluza-Klein theory and the solution thought of as simultaneously a wave and a monopole whose charge is equal to the wave's momentum.

Let us look at the moduli of the solution. The solution is specified by two pieces of data, the vector $a^M$ and the constant $h$ that appears in the harmonic function. The vector specifies the direction the wave is propagating in. That is, it gives the direction along which there is momentum. The constant $h$ in the harmonic function of the solution is then proportional to the amount of momentum carried. 

In addition, the solution comes with a monopole-like structure, whose fibre is in the direction dual to the direction of propagation of the wave and whose base is in the external spacetime. In the case of the smeared solution studied in this paper this fibration may be classified by its first Chern class which is $h$. (See \cite{bermanrudolphlocalised} for a discussion of the localized non-smeared solution.)

To give a non-trivial first Chern class the fibre must be an $S^1$ and then the {\it{magnetic charge}} $h$ is integral. This is essentially Dirac quantization but now our theory also requires self-duality which in turn implies that the momentum in the dual direction to the fibre is quantized. The presence of quantized momentum in this direction then implies that this direction itself must also be an $S^1$. Let us examine this quantitatively.

The electric charge of the solution is related to the radius of the circle by
\begin{equation}
q_e = \frac{n}{R_e}  \qquad  {\rm{with}} \, \, n \in \mathbb{Z}
\end{equation}
and the magnetic charge is related to the radius of the fibre by
\begin{equation}
q_m = m  R_m  \qquad  {\rm{with}} \, \, m \in \mathbb{Z}   \, .
\end{equation}
Now the twisted self-duality relation implies 
\begin{equation}
q_e=q_m \qquad \implies \qquad n/m =  R_e R_m \, .
\end{equation}
From examining the norms of the $E_7$ vectors that specify the solution we can determine $R_e$ and $R_m$ as
\begin{equation}
R_e=|a^M| \qandq R_m=|\ta^M| \, .
\end{equation}
Calculating these norms using the metric of the solution (see the next section) then reveals 
\begin{equation}
R_e=H^{3/4}   \, , \quad R_m= H^{-3/4}  \qquad {\rm{thus}} \qquad 
R_e R_m =1   \quad {\rm{and}} \quad  n=m.
\end{equation}
So the $E_7$ related radii are duals and the electric and magnetic quantum numbers are equal. Note that the harmonic function $H$ (and thus the radii) is a function of $r$, the radial coordinate of the external spacetime. This will lead to interesting insights when we analyze the solution close to its core or far away from it in Section \ref{sec:singularity}.

The actual direction $a^M$ that one chooses determines how one interprets the solution in terms of the various usual supergravity descriptions. That is we can interpret this single solution in terms of the brane solutions in eleven dimensions or the Type IIA and Type IIB brane solutions in ten dimensions. We will show this in detail in Section \ref{sec:spectrum}.

Finally let us a add a comment about the topological nature of these solutions. The more mathematically minded reader will note that brane solutions like the NS5-brane are not classified by the first Chern class which in cohomology terms is given by $H^2(M;\mathbb{Z})$ but instead by the Dixmier-Douady class, i.e. $H^3(M;\mathbb{Z})$. For the smeared solution these two are related since $H^3(S^2\times S^1;\mathbb{Z}) = H^2(S^2;\mathbb{Z})\times H^1(S^1;\mathbb{Z})$. Thus for the smeared branes there is no issue. The question of the global structure of the localized solutions where one has a genuine $H^3$ is however an important open question that has recently received some attention \cite{Papadopoulos:2014mxa, Papadopoulos:2014ifa, Hull:2014mxa}.

\subsection{Twisted Self-duality }
\label{sec:selfduality}
The EFT gauge potential ${\AAA_\mu}^M$ presented above satisfies the twisted self-duality equation \eqref{eq:selfduality}. This can be checked explicitly by looking at the components of the equation and making use of the relation between the harmonic function $H$ and the spacetime vector potential $A_i$ given in \eqref{eq:AH}.

First though, we will look at the relation between the two vectors $a^M$ and $\ta^M$ that define the directions of ${\AAA_t}^M$ and ${\AAA_i}^M$. The duality relation between them can be made precise by normalizing the vectors using the generalized metric $\MM_{MN}$. The unit vectors
\begin{equation}
\ha^M = \frac{a^M}{|a^N|} = \frac{a^M}{\sqrt{a^Ka^L\MM_{KL}}} \qandq
\hta^M = \frac{\ta^M}{|\ta^N|} = \frac{\ta^M}{\sqrt{\ta^K\ta^L\MM_{KL}}}
\end{equation}
are related via the symplectic form $\Omega$ by
\begin{equation}
\ha^M = \Omega^{MN}\MM_{NK}\hta^K\, , \qquad 
\Omega_{MN}=\begin{pmatrix} 0 & \mathbb{I} \\ -\mathbb{I} & 0 \end{pmatrix} \, .
\label{eq:vectorduality}
\end{equation}
If the vectors are not normalized the metric in the duality relation introduces extra factors. For the specific directions given above we have $\ha^M=H^{-3/4}\delta^{Mz}$ and $\hta^M=H^{3/4}\delta^{M\tz}$ which indeed satisfy \eqref{eq:vectorduality} for the $\MM_{MN}$ given in \eqref{eq:genmetric}.

Let's now turn to the self-duality of the field strength. We begin by computing the field strength ${\FF_{\mu\nu}}^M$ of ${\AAA_\mu}^M$ as given in \eqref{eq:genfieldstrength}, recalling the simplifications our solution provides. There are two components which read
\begin{equation}
\begin{aligned}
{\FF_{it}}^M &= 2\partial_{[i}{\AAA_{t]}}^M 
	= -\partial_i(H^{-1}-1)a^M = H^{-2}\partial_i H \delta^{Mz} \\
{\FF_{ij}}^M &= 2\partial_{[i}{\AAA_{j]}}^M 
	= 2\partial_{[i}A_{j]}\ta^M ={\epsilon_{ij}}^k\partial_k H \delta^{M\tz} \, .
\end{aligned}
\label{eq:F}
\end{equation}
The spacetime metric $g_{\mu\nu}$ is given in \eqref{eq:exmetric} and has determinant $e^2=|\det g_{\mu\nu}|=H$. This can be used to rewrite the self-duality equation \eqref{eq:selfduality} as
\begin{align}
{\FF_{\mu\nu}}^M &= \frac{1}{2}H^{1/2} \epsilon_{\mu\nu\rho\sigma}  g^{\rho\lambda}g^{\sigma\tau}\Omega^{MN} \MM_{NK}{\FF_{\lambda\tau}}^K
\end{align}
where the spacetime metric is used to lower the indices on $\FF^M$. Now we can look at the components of the equation. Starting with
\begin{equation}
{\FF_{ij}}^M = H^{1/2} \epsilon_{ijkt}  g^{kl}g^{tt}\Omega^{MN} \MM_{NK}{\FF_{lt}}^K
\end{equation}
and inserting for the spacetime metric and the field strength gives
\begin{align}
{\FF_{ij}}^M &= -H^{1/2} \epsilon_{tijk}H^{-1/2}\delta^{kl}(-H^{1/2})
		\Omega^{MN} \MM_{NK}H^{-2}\partial_l H a^K \notag\\
	&= H^{-3/2}({\epsilon_{ij}}^k\partial_k H) \Omega^{MN} \MM_{NK} \delta^{Kz}
\end{align}
where the extra minus sign in the first line comes from permuting the indices on the four-dimensional epsilon which is then turned into a three-dimensional one. In the next step we make use of \eqref{eq:AH} and the components of $\Omega$ and $\MM$ that are picked out by the summation over indices are substituted
\begin{align}
{\FF_{ij}}^M &= H^{-3/2}2\partial_{[i}A_{j]}\Omega^{Mz} \MM_{zz} \notag \\
	&= H^{-3/2}2\partial_{[i}A_{j]}\delta^{M\tz}H^{3/2} 
	= 2\partial_{[i}A_{j]}\ta^M 
\end{align}
and we obtained the expected result. Similarly, the other component of the self-duality equation reads
\begin{equation}
{\FF_{it}}^M = \frac{1}{2}H^{1/2} \epsilon_{itjk} g^{jp}g^{lq} 
		\Omega^{MN} \MM_{NK}{\FF_{pq}}^K \, .
\end{equation}
Going through the same steps as before leads to 
\begin{align}
{\FF_{it}}^M &= -\frac{1}{2}H^{1/2} \epsilon_{tijk}H^{-1/2}\delta^{kp}H^{-1/2}\delta^{lq}
		\Omega^{MN} \MM_{NK}2\partial_{[p}A_{q]}\ta^K \notag\\
	&= -H^{-1/2}({\epsilon_i}^{jk}\partial_j A_k) \Omega^{MN} \MM_{NK} \delta^{K\tz} \, .
\end{align}
Again substituting for $\Omega$ and $\MM$ gives the expected result	
\begin{align}
{\FF_{it}}^M 	&= -H^{-1/2}\partial_i H\Omega^{M\tz} \MM_{\tz\tz} \notag \\
	&= -H^{-1/2}\partial_i H(-\delta^{Mz})H^{-3/2} = H^{-2}\partial_i Ha^M 
\end{align}
to match with \eqref{eq:F}.

Thus the components of the field strength of the EFT vector ${\AAA_\mu}^M$ given in \eqref{eq:F} satisfy the self-duality condition. It is also possible to satisfy an anti-self-duality equation.  If the magnetic charge of our solution is taken to be minus the electric charge, this has the effect of modifying the magnetic component of the EFT vector by an extra minus sign, ${\AAA_i}^M=-A_i\ta^M$. The above calculation then works exactly the same but the extra minus sign ensures that the field strength is anti-self-dual. This choice would then be consistent with the original EFT paper \cite{Hohm:2013uia} (of course the choice of self-dual or anti-self-dual is ultimately related to how supersymmetry is represented).

\section{The Spectrum of Solutions}
\label{sec:spectrum}

The self-dual EFT solution presented in the previous section gives rise to the full spectrum of 1/2 BPS solutions in eleven-dimensional supergravity and the Type IIA and Type IIB theories in ten dimensions. We will now show how applying the appropriate solution to the section condition and rotating our solution in a specific direction of the exceptional extended space leads to the wave solution, the fundamental, solitonic and Dirichlet p-branes, the KK-branes which are extended monopoles, and an example of an intersecting brane solution. All these extracted solutions together with their Kaluza-Klein decomposition can be found in Appendix \ref{app:glossary} for easy referral.

\subsection{Supergravity Solutions}
We start by looking at the EFT solution from an eleven-dimensional supergravity point of view. Using the results of Section \ref{sec:embedding} in reverse, the supergravity fields can be extracted from the EFT solution. Recall that the resulting supergravity fields will be rearranged according to a $4+7$ Kaluza-Klein coordinate split. 

First, the extended coordinates $Y^M$ are decomposed into $y^m$, $y_{mn}$ and so on as given in \eqref{eq:gencoordM}. Then by comparing the expression for the generalized metric of the internal extended space, $\MM_{MN}$, of our solution in \eqref{eq:genmetric} to \eqref{eq:genmetricSUGRA}, one can work out the seven-dimensional internal metric $g_{mn}$. The components of the EFT vector potential ${\AAA_\mu}^M$ given in \eqref{eq:vecpot} can be related to the KK-vector of the decomposition and the $C_3$ and $C_6$ form fields respectively according to \eqref{eq:vecpotSUGRA}. Finally, the external spacetime metric $g_{\mu\nu}$ in \eqref{eq:exmetric} is simply carried over to the 4-sector of the KK-decompostion. 

As mentioned before, the EFT solution is characterized by the direction of the vector $a^M$ and a corresponding ordering in the diagonal entries of $\MM_{MN}$. If the procedure of extracting a supergravity solution just described is applied to the EFT solution as presented in Section \ref{sec:solution}, i.e. with the direction of the $y^m$-type, $a^M=\delta^{Mz}$ where we now identify $z$ with $y^1$, the first of the ordinary $y^m$ directions, the \emph{pp-wave} solution of supergravity can be extracted. From $\MM_{MN}$, the internal metric is given by
\begin{equation}
g_{mn} = \diag [H, \delta_6] 
\label{eq:intmetricWM}
\end{equation}
where $\delta_6$ is a Kronecker delta of dimension six. These are the remaining six directions of $y^m$. The ``electric'' part of the EFT vector, ${\AAA_t}^z = -(H^{-1}-1)$, becomes the cross-term in the supergravity metric. The ``magnetic'' part ${\AAA_i}^{\tz}=A_i$ is like a dual graviton and does not appear in the supergravity picture. Note that the dual direction to $z$ is $y_1=\tz$. See Appendix \ref{app:SUGRA} for the supergravity wave decomposed under a $4+7$ split. Since our self-dual EFT solution is interpreted as a wave now propagating in the ordinary direction $y^1=z$, it is not too surprising to recover the supergravity wave once the extra exceptional aspects are removed.

As shown in previous work \cite{Berkeley:2014nza,Berman:2014jsa}, we know that a wave in an exceptional extended geometry can also propagate along the novel dimensions such as $y_{mn}$ or $y^{mn}$. If our solution is rotated to propagate in those directions, e.g. $a^M={\delta^M}_{12}$ or $a^M=\delta^{M\ 67}$, the \emph{membrane} and \emph{fivebrane} solutions of supergravity are recovered. For the former, the membrane is stretched along $y^1$ and $y^2$, for the latter, the fivebrane is stretched along the complimentary directions to $y^6$ and $y^7$, i.e. $y^1,y^2,y^3,y^4$ and $y^5$. This result is obtained by an accompanying rotation of the generalized metric according to \eqref{eq:rot} and extracting the internal metrics for the M2 and the M5 (cf. Appendix \ref{app:SUGRA})
\begin{align}
g_{mn} &= H^{1/3}\diag [H^{-1}\delta_2, \delta_5] \, , &
g_{mn} &= H^{2/3}\diag [H^{-1}\delta_5, \delta_2] \, .
\label{eq:intmetricM2M5}
\end{align}
The masses and charges of the branes are provided by the momentum in the extended directions. The electric potential is given by ${\AAA_t}^M$ which encodes the $C_3$ for the M2 and the $C_6$ for the M5. The magnetic potential is given by ${\AAA_i}^M$ which gives their duals, i.e. the $C_6$ for the M2 and the $C_3$ for the M5. We will explain this procedure of obtaining the membrane and fivebrane from the EFT solution in more detail below.

We have previously speculated \cite{Berman:2014jsa} that the wave in EFT along $y_m$, the fourth possible direction, should correspond to a monopole-like solution in supergravity. Since we are now working with a self-dual solution, we can show that this is indeed the case. If the direction of $a^M$ is of the $y_m$-type, e.g. $a^M=\delta^{M\tz}$, and thus $\ta^M$ along $y^m$ (essentially swapping $a^M$ and $\ta^M$ of the pp-wave), the \emph{KK-monopole} is obtained. Again performing the corresponding rotation of the generalized metric, the internal metric can be extracted
\begin{equation}
g_{mn} = \diag [H^{-1}, \delta_6] \, .
\label{eq:intmetricKKM}
\end{equation}
The ``magnetic'' part of the EFT vector, ${\AAA_i}^z = A_i$, becomes part of the KK-monopole metric in supergravity. The ``electric'' part ${\AAA_t}^{\tz} = -(H^{-1}-1)$ now has the nature of a dual graviton and does not contribute in the supergravity picture. This is the opposite scenario to the pp-wave described above, underlining the electric-magnetic duality of these two solutions.

The four supergravity solutions we have extracted from our EFT solution all have the same external spacetime metric $g_{\mu\nu}$ under the KK-decomposition,
\begin{equation}
g_{\mu\nu} = \diag [-H^{-1/2},H^{1/2}\delta_{ij}]
\end{equation}
which has the character of a point-like object (in four dimensions). The four solutions only differ in the internal metric $g_{mn}$, the KK-vector of the decomposition and of course the C-fields. But these elements are just rearranged in ${\AAA_\mu}^M$ and $\MM_{MN}(g_{mn})$ and are all the same in EFT, up to an $SO(56)$ rotation of the direction $a^M$ of the solution.

\subsubsection{From Wave to Membrane}
Let's pause here briefly and take a closer look at a specific example of such a rotation. The EFT solution presented in Section \ref{sec:solution} with the choice for the vector $a^M$ given there and the generalized metric $\MM_{MN}$ in \eqref{eq:genmetric} for a fixed coordinate system directly reduces to the pp-wave in eleven dimensions. 

We now want to demonstrate how this can also give the M2-brane at the same time by simply picking a different duality frame, that is choosing a different section of the extended space to give the physical spacetime. This new duality frame is obtained by rotating the fields of the solution according to \eqref{eq:rot}. 

As explained above, if the EFT solution is propagating along a $y_{mn}$ direction, say $y_{12}$, it gives the membrane. Thus the vector $a^M=\delta^{M1}$ has to be rotated into ${a'}^M={\delta^M}_{12}$. This has the effect of exchanging $y^1$ with $y_{12}$ and their corresponding components in the metric, i.e. $\MM_{1\,1} \leftrightarrow \MM^{12\, 12}$. This should not come as a surprise since here momentum and winding directions are exchanged which is exactly what is expected in relating the wave and the membrane via duality.

Besides $y^1 \leftrightarrow y_{12}$, the frame change also swaps the following pairs of coordinates and the corresponding components of the metric (here the index $a$ takes the values 3 to 7)
\begin{equation}
\begin{aligned}
y^2 &\leftrightarrow y_2 \, , &
y_{ab} &\leftrightarrow y^{ab} \, , &
y_{1a} &\leftrightarrow y^{1a} \, .
\end{aligned}
\label{eq:pp-M2}
\end{equation}
These are simple exchanges between dual pairs of coordinates that reflect the new duality frame.

After the rotation, the generalized metric reads (still in the coordinate system given by \eqref{eq:gencoordM})
\begin{equation}
\begin{aligned}
\MM_{MN} = \diag[&H^{-1/2}\delta_2, H^{1/2}\delta_5, H^{3/2}, H^{1/2}\delta_{10}, H^{-1/2}\delta_{10}, \\
&H^{1/2}\delta_2, H^{-1/2}\delta_5, H^{-3/2}, H^{-1/2}\delta_{10}, H^{1/2}\delta_{10}] \, .
\end{aligned}
\label{eq:genmetricM2}
\end{equation}
This can now be compared to \eqref{eq:genmetricSUGRA} to read off the internal metric in the reduced, eleven-dimensional picture as described above, and gives \eqref{eq:intmetricM2M5}, the M2-brane stretched along $y^1$ and $y^2$. Similar rotation procedures can be applied to relate the pp-wave or the membrane to the fivebrane and the monopole.

Our self-dual EFT wave solution with attached monopole-structure thus unifies the four classic eleven-dimensional supergravity solutions and provides the so-far missing link in the duality web of exceptionally extended solutions.

\subsubsection{The Membrane / Fivebrane Bound State}
The self-dual EFT solution does not only give the standard 1/2 BPS branes of supergravity but also bound states. Such solutions were first mentioned in \cite{Gueven:1992hh} and then interpreted by Papadopoulos and Townsend in \cite{Papadopoulos:1996uq}. As an illustrative example we will show how the dyonic M2/M5-brane solution of \cite{Izquierdo:1995ms} can be obtained from our EFT solution. 

Before we find this bound state of a membrane and a fivebrane, it is useful to see how to pick a duality frame such that the EFT solution reduces to the (pure) fivebrane. Above we have just seen how to rotate the frame to get the (pure) membrane instead of the wave. If we rotate further to have the solution propagate in the $y^{67}$ direction, i.e ${a''}^M=\delta^{M\, 67}$, then we get the fivebrane. 

Starting from the membrane frame of the previous subsection with ${a'}^M={\delta^M}_{12}$ and the generalized metric in \eqref{eq:genmetricM2}, the new frame rotation exchanges the membrane direction $y_{12}$ with the fivebrane direction $y^{67}$ and their corresponding components in the metric, $\MM^{12\, 12} \leftrightarrow \MM_{67\, 67}$. Again it is very natural to exchange a membrane coordinate $y_{mn}$ with a fivebrane coordinate $y^{mn}$ in this kind of duality transformation.

In what follows, it will be useful to split the index of the coordinate $y^m$ into $m=(a,A,\alpha)$. Here $y^a$ with $a=1,2$ are the two worldvolume directions of the membrane or two of the worldvolume directions of the fivebrane. The $y^A$ with $A=3,4,5$ are the remaining three worldvolume directions of the fivebrane, they are transverse directions for the membrane. And finally the $y^\alpha$ with $\alpha=6,7$ are transverse directions to both the membrane and fivebrane. Table \ref{tab:coords} shows the worldvolume (a circle $\circ$) and transverse (a dash $-$) directions of the membrane and fivebrane together with the coordinate labels and sector under the KK-decomposition.
\begin{table}[h]
\centering
\begin{tabular}{|l|cccc|ccccccc|} \hline
sector		& \multicolumn{4}{c|}{external} & \multicolumn{7}{c|}{internal} \\ \hline
\multirow{2}{*}{coordinate}	 &      & \multicolumn{3}{|c|}{$w^i$} & \multicolumn{2}{c}{$y^a$} & \multicolumn{3}{|c|}{$y^A$} & \multicolumn{2}{c|}{$y^\alpha$} \\
    & \multicolumn{1}{l|}{$t$} & $w^1$ & $w^2$ & $w^3$ & $y^1$ & $y^2$ & \multicolumn{1}{|l}{$y^3$} &  $y^4$  & \multicolumn{1}{l|}{$y^5$}   & $y^6$   & $y^7$     \\ \hline
membrane    & $\circ$   &  -    &  -    &  -    & $\circ$     & $\circ$     & -      & -   &   - & -   & - \\
fivebrane    & $\circ$   & -     &  -    &  -    & $\circ$     & $\circ$  & $\circ$   &  $\circ$  & $\circ$  & -   & -    \\ \hline
\end{tabular}
\caption{The coordinates of the membrane and fivebrane are either in the external or internal sector of the KK-decomposition. A circle $\circ$ denotes a worldvolume direction of the brane while a dash $-$ indicates a transverse direction.}
\label{tab:coords}
\end{table}

Besides $y_{ab}=y_{12} \leftrightarrow y^{\alpha\beta}=y^{67}$, the frame change also swaps some other pairs of coordinates. This can now be neatly written as 
\begin{equation}
\begin{aligned}
y^{ab} &\leftrightarrow y_{\alpha\beta} \, , &
y_{a\alpha} &\leftrightarrow y^{a\alpha} \, , \\
y^A &\leftrightarrow y_{BC} \, , &
y_A &\leftrightarrow y^{BC} \, .
\end{aligned}
\label{eq:M2-M5}
\end{equation}
The first line contains further exchanges between membrane directions and fivebrane directions. The second line is a result of going from the membrane frame to the fivebrane frame. Once the corresponding components of the metric have been exchanged as well, it reads
\begin{equation}
\begin{aligned}
\MM_{MN} = \diag[&H^{-1/2}\delta_5, H^{1/2}\delta_2, H^{1/2}\delta_{10}, H^{-1/2}\delta_{10}, H^{-3/2}, \\
&H^{1/2}\delta_5, H^{-1/2}\delta_2, H^{-1/2}\delta_{10}, H^{1/2}\delta_{10},  H^{3/2}]\, .
\end{aligned}
\label{eq:genmetricM5}
\end{equation}
which can be reduced to give the internal fivebrane metric \eqref{eq:intmetricM2M5}. Inserting the rotated vector ${a''}^M$ into \eqref{eq:vecpot} then gives the corresponding C-form field as explained above.

Now that it is clear how to obtain both the M2-brane and the M5-brane from our self-dual EFT solution, we can attempt to obtain the dyonic M2/M5 bound state of \cite{Izquierdo:1995ms}. To achieve this, we will again start from the membrane duality frame. This time though, we do not rotate the frame all the way into the fivebrane frame but introduce a parameter $\xi$ which interpolates between a purely electric M2-brane and a purely magnetic M5-brane. For $\xi=0$ the transformation gives the fivebrane whereas for $\xi=\pi/2$ the membrane is recovered\footnote{This choice of $\xi$ -- and not one shifted by $\pi/2$ -- might be counter-intuitive but has been made to match the $\xi$ in \cite{Izquierdo:1995ms}.}. Therefore a vector of the form
\begin{equation}
a^M_{\mathrm{(M2/M5)}} = \sin\xi \, a^M_{\mathrm{(M2)}} + \cos\xi \, a^M_{\mathrm{(M5)}} \\
\label{eq:aM2M5}
\end{equation}
points the EFT solution in the direction which gives the M2/M5-brane (here $a^M_{\mathrm{(M2)}}={\delta^M}_{12}$ and $a^M_{\mathrm{(M5)}}=\delta^{M\, 67}$ from above). If this vector is inserted into the EFT vector potential, one obtains both the $C_3$ and the $C_6$ (together with their duals) of the membrane and fivebrane, each modulated by $\sin\xi$ or $\cos\xi$. Since we are dealing with a dyonic solution, both an electric and a magnetic potential are expected. 

Having found the new EFT vector, the above rotation now needs to be applied to the generalized metric. Comparing the metric for the M2 in \eqref{eq:genmetricM2} and the M5 in \eqref{eq:genmetricM5}, one finds that the components which get exchanged in \eqref{eq:M2-M5} differ by a factor of $H$. In most cases $H^{1/2}$ becomes $H^{-1/2}$ or vice versa, e.g. $\MM_{A\, B} = H^{1/2}\delta_{AB}$ and $\MM^{AB\, CD} = H^{-1/2}\delta^{AB,CD}$ are exchanged. The only exceptions are for the $ab=12$ and $\alpha\beta=67$ components where $H^{\pm 3/2}$ becomes $H^{\pm 1/2}$. The partial, $\xi$-dependent rotation now introduces factors of $\sin\xi$ and $\cos\xi$ into the metric components and generates off-diagonal entries. To see how they arise, one has to consider the effect of the rotation on the coordinates.

The coordinate pairs which get rotated into each other are the same as in \eqref{eq:M2-M5}, but now superpositions are formed instead of exchanging them completely. One can think of each pair as a 2-vector acted on by 
\begin{equation}
R_2=\begin{pmatrix}\sin\xi & \cos\xi \\ -\cos\xi & \sin\xi \end{pmatrix}
\end{equation}
which is the $2\times 2$ submatrix of the full rotation matrix $R$ in \eqref{eq:rot}. Then the new coordinate pair is schematically\footnote{More formally, one can introduce epsilon symbols so that the index structure works out, e.g. ${y'}^A = \sin\xi\, y^A + \cos\xi\ \epsilon^{ABC}y_{BC}$.} given by, for example 
\begin{equation}
\begin{pmatrix}
{y'}^A \\ y'_{BC}
\end{pmatrix} =
R_2 \begin{pmatrix}
y^A \\ y_{BC}
\end{pmatrix} =  
\begin{pmatrix}
\sin\xi\ y^A + \cos\xi\ y_{BC} \\ 
\sin\xi\ y_{BC} - \cos\xi\ y^A
\end{pmatrix} \, .
\label{eq:rotationbyxi}
\end{equation}
The other coordinate pairs which are acted on by copies of $R_2$ are
\begin{equation}
\begin{pmatrix}
y_{ab} \\ y^{\alpha\beta}
\end{pmatrix} , \ 
\begin{pmatrix}
y^{ab} \\ y_{\alpha\beta}
\end{pmatrix} , \
\begin{pmatrix}
y_{a\alpha} \\ y^{b\beta}
\end{pmatrix} \quad\mathrm{and}\quad
\begin{pmatrix}
y_A \\ y^{BC}
\end{pmatrix} .
\label{eq:coordsM2M5}
\end{equation}
These rotations have quite non-trivial consequences for the corresponding components of the generalized metric. Conjugating the $2\times 2$ blocks of the metric with $R_2$ gives for our example
\begin{equation}
\begin{pmatrix}
\MM'_{A\, B} & {\MM'_A}^{EF} \\ {{\MM'}^{CD}}_{B} & {\MM'}^{CD\, EF}
\end{pmatrix} = 
R_2
\begin{pmatrix}
\MM_{A\, B} & 0 \\ 0 & \MM^{CD\, EF}
\end{pmatrix}
R_2^{\, -1}
\end{equation}
and similarly for all the other metric components which are rotated into each other. The essential action of this rotation becomes clearest when the indices are suppressed. The result, which is the same for all the blocks, is
\begin{equation}
R_2
\begin{pmatrix}
H^{1/2} & 0 \\ 0 & H^{-1/2}
\end{pmatrix}
R_2^{\, -1} =
\begin{pmatrix}
H^{1/2}\sin^2\xi + H^{-1/2}\cos^2\xi & -H^{-1/2}(H-1)\sin\xi\cos\xi \\ 
-H^{-1/2}(H-1)\sin\xi\cos\xi & H^{-1/2}\sin^2\xi + H^{1/2}\cos^2\xi
\end{pmatrix} \, .
\end{equation}
This transformation produces additional off-diagonal terms in the generalized metric. In the ordinary supergravity picture these extra terms reduce to components of the C-field in the internal sector of the KK-decomposition which are of the form $C_{mnk}$. These terms are not present for the pure membrane and fivebrane, they only occur in the bound state solution. The $E_7$ generalized metric with cross-terms due to non-vanishing internal C-field was constructed in general form in \cite{Berman:2011jh} and the appropriate reduction ansatz (which has the same form as the metric) for our concrete scenario was spelled out in \cite{Berman:2014jsa}. 

The next step is thus to bring the above matrix into the standard coset form of a generalized metric or a KK-reduction ansatz which can be done by using some trigonometric identities and introducing the shorthand
\begin{equation}
\Xi = \sin^2\xi + H \cos^2\xi \, .
\label{eq:Xi}
\end{equation}
Then the new metric components read
\begin{equation}
\begin{pmatrix}
H^{1/2}\Xi^{-1}\left[1+\frac{(H-1)^2}{H}\sin^2\xi\cos^2\xi\right] & 
-H^{-1/2}\Xi \frac{H-1}{\Xi}\sin\xi\cos\xi \\ 
-H^{-1/2}\Xi \frac{H-1}{\Xi}\sin\xi\cos\xi &
H^{-1/2}\Xi 
\end{pmatrix} 
\end{equation}
which is of the desired form. It is interesting to see that it is actually possible to rewrite the transformed metric in a coset form. The underlying reason for this is that the original matrix was already in coset form, just without any off-diagonal terms, i.e. without a C-field. 

Now this matrix can be compared to a suitable reduction ansatz to extract the (components of) the metric and the C-field in supergravity. Such an ansatz -- adapted to our coordinates here -- takes the form
\begin{equation}
g^{1/2}
\begin{pmatrix}
g_{AB} + C_{ACD}g^{CD,EF}C_{EFB} & C_{ACD}g^{CD,EF} \\
g^{CD,EF}C_{EFB} & g^{CD,EF}
\end{pmatrix}
\end{equation}
where $g = \det g_{mn}$ is the determinant of the internal metric. Comparing these two matrices leads to the following components (note that of course to find the determinant all blocks have to be taken into account, not just those corresponding to $y^A$ and $y_{BC}$)
\begin{equation}
\begin{aligned}
g_{AB} &= H^{1/3}\Xi^{-2/3} \delta_{AB} & 
g &= H^{1/3}\Xi^{-2/3}  \\
g^{CD,EF} &= H^{-2/3}\Xi^{4/3} \delta^{CD,EF} &
C_{ABC} &= -\frac{H-1}{\Xi}\sin\xi\cos\xi\ \epsilon_{ABC}\, .
\end{aligned}
\end{equation}

The same procedure also works for the other pairs of indices that need to be transformed and their corresponding metric components. The only difference is for $(y_{12}, y^{67})$ where the metric has an extra factor of $H$, i.e. $\MM^{12,12} = H^{3/2}$ and $\MM_{67,67} = H^{1/2}$. But this factor is just carried through and does not affect the calculation presented above. Similarly, for $(y^{12}, y_{67})$ there is an extra factor of $H^{-1}$ in the metric.

Also note that for some cross-terms in the reduction ansatz one needs to define $V^{m_1\dots m_4} = \frac{1}{3!}\epsilon^{m_1\dots m_4n_1\dots n_3}C_{n_1\dots n_3}$, see \cite{Berman:2011jh} for more details. Since the only non-zero component (in the internal sector) of $C_3$ is $C_{ABC}$, the only component of this $V$ that does not vanish is $V^{ab\alpha\beta}=-\frac{H-1}{\Xi}\sin\xi\cos\xi\epsilon^{ab\alpha\beta}$.

Once the transformation of each index pair and the corresponding metric component together with the reduction to supergravity is performed, the dyonic M2/M5-brane solution 
is obtained in the usual 4+7 Kaluza-Klein split. Its internal metric $g_{mn}$ (and its determinant $g$) recovered form the generalized metric together with the external metric $g_{\mu\nu}$ which is just carried over from the external sector of the EFT solution are given by
\begin{equation}
\begin{aligned}
g_{mn} &= H^{1/3}\Xi^{1/3}\diag[H^{-1}\delta_{ab},\Xi^{-1}\delta_{AB},\delta_{\alpha\beta}] \, , & 
g &=H^{1/3}\Xi^{-2/3}  \\
g_{\mu\nu} &= \diag[-H^{-1/2},H^{1/2}\delta_{ij}]  \, .
\end{aligned}
\end{equation}
with $\Xi$ as defined in \eqref{eq:Xi}. Reversing the KK-decomposition, finally gives the eleven-dimensional spacetime metric of the solution as in \cite{Izquierdo:1995ms}
\begin{equation}
\begin{aligned}
\dd s^2 &= H^{-2/3}\Xi^{1/3}[-\dd t^2 + \delta_{ab}\dd y^a \dd y^b]
			+ H^{1/3}\Xi^{-2/3}[\delta_{AB}\dd y^A \dd y^B] \\
		&\qquad	+ H^{1/3}\Xi^{1/3}[\delta_{ij}\dd w^i \dd w^j 
			+ \delta_{\alpha\beta}\dd y^\alpha \dd y^\beta] \, .
\end{aligned}
\end{equation}
The harmonic function $H=H(r)$ here only depends on the three $w^i$ where $r^2=\delta_{ij}w^iw^j$. It is smeared over the remaining transverse coordinates $y^A$ and $y^\alpha$. In \cite{Izquierdo:1995ms} the solution is only delocalized in the three $y^A$ since it is constructed in eight dimensions and then lifted to eleven dimensions by including the $y^A$. Simply delocalizing it in $y^\alpha$ allows for a complete identification with the solution here. Furthermore, in the reference a multi-brane solution is constructed whereas here only a single source is considered. The result can of course be extended to take several identical brane sources into account.

It can be checked that setting $\xi$ to $0$ or $\pi/2$ and thus either $\Xi=H$ or $\Xi=1$ reproduces the pure M5-brane \eqref{eq:M5} and the pure M2-brane \eqref{eq:M2} respectively.

The components of the three-form gauge potential which have one external and two internal indices, i.e. $C_{\mu\ mn}$ were obtained from the EFT vector potential ${\AAA_\mu}^M$. The component $C_{mnk}$ which is entirely in the internal sector was extracted from the generalized metric $\MM_{MN}$. Together they read
\begin{equation}
\begin{aligned}
C_{tab} &= \frac{H-1}{H}\sin\xi\ \epsilon_{ab} \\
C_{i\alpha\beta} &= A_i\cos\xi\ \epsilon_{\alpha\beta} \\
C_{ABC} &= -\frac{H-1}{\Xi}\sin\xi\cos\xi\ \epsilon_{ABC} 
\end{aligned}
\label{eq:M2M5C3}
\end{equation}
where $A_i$ is defined as before. These are exactly the C-field components of the bound state solution (in \cite{Izquierdo:1995ms} they are given in terms of their field strengths). Again one can check that in the pure cases where either $\cos\xi=0$ or $\sin\xi=0$, the three-form potential only has a single component as given in \eqref{eq:M2} and \eqref{eq:M5} respectively. The third component above vanishes in the two pure cases. 

An interesting observation is that $\Xi=\sin^2\xi + H \cos^2\xi$ goes to $1$ far away from the brane solution since $H\rightarrow 1$ for $r\rightarrow\infty$ while near the core where $\Xi\sim H$ the fivebrane geometry prevails.

In summary, it has been shown that the self-dual EFT solution contains the dyonic M2/M5-brane solution. Therefore, the EFT solution does not only give the standard supergravity branes but in fact also the brane bound states. The standard ones are the objects obtained by pointing the vector along one of the axes of our 56-dimensional exceptional extended coordinate space. But any combination of directions is possible, thus giving rise to dyonic bound states of branes. Furthermore, the solutions of the Type II theories in ten dimensions are also included. We will look at this aspect next. 

\subsection{Type IIA Solutions}
In Appendix \ref{app:embedding} it is shown how the ten-dimensional Type IIA theory can directly be embedded into EFT without an intermediate step to the eleven-dimensional theory. Applying this procedure in reverse, the EFT solution can be viewed from a Type IIA point of view.

In the case of extracting the eleven-dimensional solutions, the internal extended coordinate $Y^M$ was decomposed into four distinct subsets \eqref{eq:gencoordM} such as $y^m$ or $y_{mn}$. Having the EFT wave propagating along those four kinds of directions gave rise to the four different solutions in supergravity with the four components of the EFT vector potential \eqref{eq:vecpotSUGRA} providing the KK-vector and C-fields. Now in the ten-dimensional Type IIA case, the generalized coordinate splits into eight separate sets of directions \eqref{eq:gencoordIIA} and we can thus expect to get eight different solutions, one for each possible orientation of the EFT solution (together with the eight types of components of the EFT vector \eqref{eq:vecpotIIA}). 

Let us first obtain the WA-solution, the pp-wave spacetime in Type IIA. The generalized metric has to be slightly reshuffled to accommodate our new choice of coordinates, its precise form can be found in the appendix. To obtain the wave, the EFT solution is made to propagate along one of the ordinary directions $y^\bm$, say $y^1=z$. Using the ansatz \eqref{eq:genmetricIIA} for $\MM_{MN}$ in Type IIA and comparing it to the (rotated) generalized metric of the EFT solution gives the dilaton $e^{2\phi}$ and the internal 6-metric $\bg_{\bm\bn}$ of the WA-solution under the $4+6$ KK-decomposition. The corresponding EFT vector component is ${\AAA_t}^z=-(H^{-1}-1)$ which provides the KK-vector of the decomposition and combines with the internal and external metrics to form the ten-dimensional metric of the wave. The other component of the vector potential, ${\AAA_i}^\tz = A_i$ is related to the dual graviton which does not appear in the ten-dimensional picture. The KK-decomposition of the wave and other Type II solutions can be found in Appendix \ref{app:Strings}.

If instead the EFT solution is chosen to propagate along the compact circle $y^\theta$, the same procedure as above leads to the D0-brane. The RR-one-form $C_1$ it couples to can be extracted from the EFT vector ${\AAA_t}^\theta$. The dual seven-form $C_7$ is derived from the other component, ${\AAA_i}^\ttheta$.

The picture should be clear by now. The EFT solution, that is the generalized metric $\MM_{MN}$ and the vector potential ${\AAA_\mu}^M$, are rotated in a specific direction. Depending on the nature of that direction, different solution in the ten-dimensional theory arise. The F1-string and NS5-brane solution can be extracted if the EFT solution propagates along one of the $y_{\bm\theta}$ and $y^{\bm\theta}$ directions respectively. The corresponding EFT vector provides the NSNS-two-form $B_2$ and dual NSNS-six-form $B_6$ for the string and vice versa for the fivebrane. Similarly, if the directions are $y_{\bm\bn}$ and $y^{\bm\bn}$, the D2- and D4-branes with the corresponding set of dual RR-three-form $C_3$ and  RR-five-form $C_5$ are obtained. 

The last two directions the EFT solution can be along are $y_\bm$ and $y_\theta$. These are the dual directions to $y^\bm$ and $y^\theta$ and hence provide the solutions dual to WA and D0, that is the KK6A-brane and the D6-brane. For the KK6A-brane, essentially the KK-monopole of the Type IIA theory, if we choose $y_1=\tz$ as the direction, the EFT vector ${\AAA_i}^\tz=A_i$ gives the KK-vector for the ten-dimensional metric and the dual ${\AAA_t}^z=-(H^{-1}-1)$ is the dual graviton for that solution. For the D6-brane, the EFT vector provides the RR-seven-form $C_7$ it couples to together with the dual one-form $C_1$.

We have thus outlined how eight different Type IIA solutions can all be extracted from a single self-dual solution in EFT. The fundamental wave and string, the solitonic monopole and fivebrane, and the four p-even D-branes all arise naturally by applying the Type IIA solution to the section condition to the EFT wave rotated in the appropriate direction. A summery of all the possible orientations and corresponding solutions can be found at the end of this section.

\subsection{Type IIB Solutions}
Along the same lines as above, using the ansatz for embedding the Type IIB theory into EFT allows for further solutions to be extracted from the EFT wave. The generalized coordinate $Y^M$ is now split into five distinct sets according to \eqref{eq:gencoordIIB} which gives five possible directions to align the EFT solution (together with five types of components in the EFT vector \eqref{eq:vecpotIIB}). 

As before, the entries of the generalized metric $\MM_{MN}$ have to be rearranged to accommodate the choice of coordinates (see Appendix \ref{app:Strings}). Comparing the Type IIB ansatz for $\MM_{MN}$ in \eqref{eq:genmetricIIB} to the (rotated) generalized metric leads to the six-dimensional internal metric $\bg_{\bm\bn}$ together with the $SL(2)$ matrix $\gamma_{ab}$. If the direction of choice is of the $y^\bm$ type, the WB-solution can be extracted. This is the pp-spacetime of the Type IIB theory which is identical to the WA-solution. The procedure is exactly the same as before with the ${\AAA_\mu}^M$ providing the KK-vector for the ten-dimensional metric (and the dual graviton which plays no role).

The dual choice of direction, i.e. $y_\bm$, gives the dual solution, that is the KK6B-brane, the KK-monopole of the Type IIB theory. Again the EFT vector contributes the KK-vector and dual graviton. The KK6B-brane is identical to the KK6A-brane.

A more interesting choice of direction is to rotate the EFT solution along one of the $y_{\bm\ a}$. This produces the Type IIB S-duality doublet of the F1-string and D1-brane. They couple to a two-form which carries an additional $SL(2)$ index $a$ to distinguish between the NSNS-field $B_2$ and the RR-field $C_2$. From the generalized metric $\MM_{MN}$ the internal metric $\bg_{\bm\bn}$ and the $SL(2)$ matrix $\gamma_{ab}$ containing the dilaton $e^{2\phi}$ ($C_0$ vanishes for this solution) can be extracted. The EFT vector $\AAA_{\mu\ \bm\ a}$ provides the two-form (and also the dual six-form).

Similarly, the EFT solution along one of the $y^{\bm\ a}$ gives rise to the other S-duality doublet of the Type IIB theory, the NS5-brane and the D5-brane. They couple to a six-form  which also carries an $SL(2)$ index to distinguish the NSNS- and RR-part, $B_6$ and $C_6$ respectively. The six-form is encoded in the electric part of the EFT vector ${\AAA_t}^{\bm\ a}=-(H^{-1}-1)$ (and the dual two-form is encoded in the magnetic part $\AAA_{i\ \bm\ a}=A_i$) upon dualization on the internal coordinate. 

Finally, having the EFT solution along the fifth direction from a Type IIB point of view, $y_{\bm\bn\bk}$, leads to the self-dual D3-brane together with its self-dual four-form $C_4$ encoded in $\AAA_{\mu\ \bm\bn\bk}$. 

As in the Type IIA theory, the fundamental wave and string, the solitonic monopole and fivebrane, and three p-odd D-branes, can all be extracted from the EFT solution by applying the Type IIB solution to the section condition and rotating the fields appropriately. All the obtained solutions are summarized in Table \ref{tab:solutions}, together with the orientation the EFT solution, i.e. its direction of propagation.

\begin{table}[h]
\centering
\begin{tabular}{|c|c|clcl|cc|}
\hline
theory & solution & \multicolumn{2}{c}{orientation}
 & \multicolumn{2}{c|}{\begin{tabular}[c]{@{}c@{}}EFT\\vector\end{tabular}} 
 & ${\AAA_t}^M$ & ${\AAA_i}^M$ \\ \hline
\multirow{5}{*}{$D=11$} 
 & WM 	&\ \ \ & $y^m$ 		&& ${\AAA_\mu}^m$ 	& KK-vector 		& dual graviton 	\\
 & M2 	&& $y_{mn}$ 		&& $\AAA_{\mu\ mn}$ 		& $C_3$ 			& $C_6$ 			\\
 & M2/M5 && 	\ *	&&		\ *	&  $C_3\oplus C_6$ 	&		$C_6\oplus C_3$		\\
 & M5 	&& $y^{mn}$ 		&& ${\AAA_\mu}^{mn}$ 	& $C_6$ 			& $C_3$ 			\\
 & KK7 	&& $y_m$ 		&& $\AAA_{\mu\ m}$ 		& dual graviton 	& KK-vector 		\\ \hline
\multirow{8}{*}{\begin{tabular}[c]{@{}c@{}}$D=10$\\Type IIA\end{tabular}} 
 & WA 	&& $y^\bm$ 			&& ${\AAA_\mu}^\bm$			& KK-vector & dual graviton 	\\
 & D0 	&& $y^\theta$ 		&& ${\AAA_\mu}^\theta$		& $C_1$ 			& $C_7$ 		\\
 & D2 	&& $y_{\bm\bn}$ 		&& $\AAA_{\mu\ \bm\bn}$		& $C_3$ 			& $C_5$ 		\\
 & F1 	&& $y_{\bm\theta}$ 	&& $\AAA_{\mu\ \bm\theta}$	& $B_2$ 			& $B_6$ 		\\
 & KK6A && $y_\bm$ 			&& $\AAA_{\mu\ \bm}$			& dual graviton 	& KK-vector 	\\ 
 & D6 	&& $y_\theta$ 		&& $\AAA_{\mu\ \theta}$		& $C_7$ 			& $C_1$ 		\\
 & D4 	&& $y^{\bm\bn}$		&& ${\AAA_\mu}^{\bm\bn}$		& $C_5$ 			& $C_3$ 		\\
 & NS5 	&& $y^{\bm\btheta}$	&& ${\AAA_\mu}^{\bm\theta}$	& $B_6$ 			& $B_2$ 		\\\hline
\multirow{5}{*}{\begin{tabular}[c]{@{}c@{}}$D=10$\\ Type IIB\end{tabular}} 
 & WB 		&& $y^\bm$ 			&& ${\AAA_\mu}^\bm$ 			& KK-vector 		& dual graviton \\
 & F1 / D1 	&& $y_{\bm\ a}$ 		&& $\AAA_{\mu\ \bm \ a}$ 		& $B_2$ / $C_2$ 	& $B_6$ / $C_6$ 	\\
 & D3 		&& $y_{\bm\bn\bk}$ 	&& $\AAA_{\mu\ \bm\bn\bk}$ 	& $C_4$ 			& $C_4$ 			\\
 & NS5 / D5 	&& $y^{\bm\ a}$ 		&& ${\AAA_\mu}^{\bm\ a}$		& $B_6$ / $C_6$ 	& $B_2$ / $C_2$ 	\\
 & KK6B 		&& $y_\bm$			&& $\AAA_{\mu\ \bm}$			& dual graviton 	& KK-vector 		\\ 
\hline
\end{tabular}
\caption{This table shows all supergravity solutions in ten and eleven dimensions discussed in this section. The orientation indicates the type of direction along which the EFT solution propagates to give rise to each of the supergravity solutions. It also determines the nature of the components of the EFT vector in the supergravity picture. * The orientation of the M2/M5 bound state requires a superposition of a membrane direction $y_{mn}$ and a fivebrane direction $y^{mn}$. Therefore the EFT vector in that hybrid direction gives both the $C_3$ and the $C_6$ since it is a dyonic solution.}
\label{tab:solutions}
\end{table}

In theory it should also be possible to obtain the D-instanton (the D(-1)-brane) and its dual, the D7-brane, from the EFT solution. The reason why this is not as straightforward as for all the other D-branes is that the instanton, as the name implies, does not have a time direction, it is a ten-dimensional Euclidean solution. Therefore the EFT solution has to be set up in such a way that the time coordinate is not in the external sector but in the internal sector of the KK-decomposition. Then being part of the exceptional extended space it can be rotated and ``removed'' when taking the section back to the physical space, leaving a solution without a time direction.

The issue for the D7-brane is that it only has two transverse directions, so it cannot fully be accommodated by our KK-decomposition which places time plus \emph{three} transverse direction in the external sector and the the world volume (with the remaining transverse bits, if there are any) in the internal sector. This clearly does not work for the D7-brane.

Both of these reasons are not fundamental shortcomings of the EFT solution, they are just technical issues arising from the way we set everything up.

\section{Singularities}
\label{sec:singularity}

Here we wish to enter into some speculation that has motivated some of this work. In particular we wish to comment on the issue of the singularity structure of supergravity solutions\footnote{We are grateful to Michael Duff for discussions on this issue.}. 

Having constructed this self-dual wave solution with monopole-structure in EFT and shown how it relates to the known solutions of supergravity, we can analyse it further. The fields of the solution, that is the metrics $g_{\mu\nu}$ and $\MM_{MN}$ and the vector potential ${\AAA_\mu}^M$, are all expressed in terms of the harmonic function $H(r)$ with $r^2=\delta_{ij}w^iw^j$ where $r$ is the radial coordinate of the transverse directions in the external sector. This leads to the immediate question of what happens to the solution when $r$ goes to zero or infinity, i.e. what happens close to the core of the solution or far away from it? 

As is well known from the works of Duff and others \cite{Duff:1991pe,Duff:1994an}, the nature of singularities at the core of brane solutions depends on the duality frame that one uses. Given that EFT provides a formalism unifying different duality frames then one would imagine that solutions in EFT maybe be singularity free at the core. Since the solution lives in an extended space, the additional dimensions help to smooth the singularity at the core. This is not the case for the fundamental string, for example.

A good way to think about DFT or EFT is as a Kaluza-Klein-type theory. The space is extended and the reduction of the theory through use of the section condition gives supergravity. Let's recall some basic properties of ordinary Kaluza-Klein theory that will be useful for our intuition. The reduced theory is gravity plus electromagnetism (and a scalar field which will not be relevant here). One typically allows various singular solutions such as electric sources which have delta function-type singularities and magnetic sources which also are singular. Then the Kaluza-Klein lift of these solutions smooths out the singularities. The electric charges are just waves propagating around the KK-circle and the magnetic charges come from fibering the KK-circle to produce a total space describing an $S^3$. Thus the singularities inherent from the abelian charges become removed when one considers the full theory and the $U(1)$ is just a subgroup of some bigger non-abelian group, in this case five-dimensional diffeomorphisms. 

A similar process happens when one considers the 't\,Hooft-Polyakov monopole where in the low energy effective field theory the gauge group is broken to $U(1)$ and the monopole is a normal Dirac monopole (with a singularity at the origin). Near the core of the monopole however, the low energy effective description breaks down, and the full non-abelian theory becomes relevant. The non-abelian interactions smooth out the core of the monopole and the singularity is removed. This intuition is exactly what we wish to envoke when thinking about DFT and EFT. Solutions become smoothed out by the embedding in a bigger theory, $U(1)$ charges in particular are simply the result of some reduction and the singularities are non-existent in the full theory \cite{Harvey:1996ur}.

So can one show that the EFT solutions described here are free of singularities at their core? 
To see how this works, we will first return to a simpler example, the wave in Double Field Theory that was presented in \cite{Berkeley:2014nza} which gives the fundamental string when the direction of propagation is a winding direction. After this short digression we will return to the EFT solution and find a similar result.

\subsection{The Core of the DFT Wave}
It is well known that the fundamental string of string theory has a singularity at its core (essentially there are delta function sources required by the solution). The existence of this singularity can easily be inferred from looking at the Ricci scalar or sending a probe towards the core of the string and looking at the proper time it takes the probe to do so \cite{Duff:1991pe,Duff:1994an}.

The other fundamental object in string theory, the T-dual of the string, is the wave. The wave is clearly non-singular as a straightforward calculation of the Riemann curvature shows. If one takes a closer look at the string and the wave, it is the cross-terms in the wave metric that ensure that the curvature remains finite and does not develop a singularity. The T-duality (essentially the Buscher rules) that turns the wave into the string moves these cross-terms into the B-field of the string. The curvature of the string then becomes singular as $r$ goes to zero. It seems that T-duality has some power over the behaviour of the solution close to the core. 

In Double Field Theory T-duality simply is a rotation in the doubled space. In DFT there is a single fundamental solution, the DFT wave. Depending on its orientation in the doubled space, either the string or the wave solution of string theory can be extracted when seen from an un-doubled point of view.

The DFT solution is non-singular everywhere. The notion of curvature in DFT is a slightly ambiguous concept. By non-singular we mean that the generalized Ricci tensor defined by varying the DFT action with respect to the generalised metric vanishes everywhere for the DFT wave. Of course the equations of motion for DFT dictate that this must vanish in the absence of any RR or fermionic sources since the NSNS sector is contained in the generalized geometry. What is significant is that one might have allowed delta function sources as one does for the Schwarzschild solution in ordinary relativity. None are required by the wave solution in DFT. The lack of a singularity at the core may also be argued from analogy with ordinary relativity. The solution in question is the wave solution for DFT and the wave solution in ordinary GR is free from singularities therefore we expect the DFT wave to also be singularity free. Thus looking at the solution from the perspective of the doubled space eliminates the singularity at the core of the fundamental string. 

One can ask further how the solution behaves closes to the core or far away, i.e. what happens when $r$ is small or large? Is there is a natural choice of picking the coordinates that form the physical spacetime? To answer this question we have to take a closer look at the DFT wave. The generalized metric of this solution can be written in terms of a line element as
\begin{align}
\dd s^2 &= \HH_{MN}\dd X^M \dd X^N \notag \\ 
		&=(H-2)\left[\dd t^2 - \dd z^2 \right] 
			+ 2(H-1) \left[\dd t\dd \tz + \dd\ttt\dd z\right]
			- H \left[\dd\ttt^2 - \dd\tz^2 \right] \notag \\
		&\qquad	+ \delta_{mn}\dd y^m\dd y^n
			+ \delta^{\bm\bn}\dd\ty_\bm\dd\ty_\bn
\end{align}
where the doubled coordinates are $X^M=(t,z,\ttt,\tz,y^m,\ty_\bm)$. The doubled space has dimension $2D$ and the transverse coordinates are labelled by $m,\bm=1,\dots,D-2$. The harmonic function here (for $D>4$) is given by $H = 1 + \frac{h}{r^{D-4}}$ with $r^2=\delta_{mn}y^my^n$. In order to study the behaviour of this solution we will take all the relevant doubled coordinates to be compact such that $(t,z,\ttt,\tz)$ are periodic coordinates.

In order to use our Kaluza-Klein inspired intuition we will first examine the behaviour of DFT on a simple $2D$-dimensional torus. The $2D$ DFT torus is decomposed into a $D$-torus with volume $R^D$ and a dual torus with volume $1/R^D$ (the volume of the DFT space is 1 as it must be). This decomposition is always possible due to the presence of the invariant $O(D,D)$ tensor (usually denoted by $\eta$) that provides the doubled space with a polarization. Double Field Theory comes equipped with a coupling 
\beq
G_{DFT}= e^{2d} \,  \alpha'^{D-1} 
\eeq
where $e^{2d}$ is the DFT dilaton and acts as a dimensionless coupling for DFT in the same way as the usual dilaton does for supergravity. The DFT coupling $G_{DFT}$ is to be compared with the usual Newton's constant in $D$-dimensional supergravity given by
\beq
G_N=e^{2\phi} \alpha'^{\frac{D-2}{2}} \, .
\eeq
It is known (almost by construction) that reducing DFT on the \textit{dual} $D$-torus \`a la Kaluza-Klein gives supergravity in $D$ dimensions and thus we can relate the coupling for the reduced theory to the DFT coupling
\beq
e^{2\phi}= \frac{R^D}{\alpha'^{D/2}} e^{2d} \, .
\eeq
Equivalently, we may instead reduce on the $D$-torus, to give the T-dual supergravity picture which gives the relation
\beq
e^{2\phi_{{\rm dual}}}=\frac{\alpha'^{D/2}}{R^D} e^{2d} \, .
\eeq
These three couplings then potentially provide a hierarchy that is governed by $R^D$, the volume of the torus. This analysis gives the completely intuitive result that for $R$ much larger than the string scale, the appropriate description is DFT reduced on the dual torus since its coupling, $e^{2\phi}$, is greatest. For small $R$ the appropriate description is for the theory reduced on the torus itself, as its coupling, $e^{2\phi_{{\rm dual}}}$, is greatest. For a circle whose radius is of order 1 there is no preferred reduction and the hierarchy breaks down. Thus the total doubled space should be taken into account without reduction. This is as it should be, for tori near the string scale we need to include both ordinary modes and the winding modes simultaneously. This somewhat pedestrian analysis is just slightly formalising the notion that on a compact space there is a natural T-duality frame that is picked out by the one where the volume is largest.{\footnote{The reader may prefer that this argument should be expressed in terms of energy scales in which case simply convert the hierarchy of gravitational couplings to effective Planck masses and examine the theory that dominates the low energy effective action.}} 

 We now wish to apply this lesson to our DFT wave solution and determine which dimensions become {\it{large}} and thus pick the {\it{T-duality frame}}. These large dimensions will be those that then become identified with the spacetime of supergravity. This is non-trivial in the sense that the volume of the compact space will be a function of $r$, the distance from the core of the solution. One should think of the solution as a toroidal fibration with a one-dimensional base space with coordinate $r$ and the toroidal fibre being given by the generalized metric. For the solution at hand this is space described by the $4\times 4$ generalized metric for the coordinates $X^A=(t,z,\ttt,\tz)$
\begin{equation}
\HH_{AB} = \begin{pmatrix}
H-2		& 0		& 0   		& H-1 	\\
0		& 2-H	& H-1   	& 0 		\\
0		& H-1	& -H	  	& 0		\\
H-1		& 0		& 0 		& H
\end{pmatrix} \, .
\end{equation}
The function $H = 1 + \frac{h}{r^{D-4}}$ completely determines the geometry and is solely a function of $r$, the radial distance to the solution's core. In these coordinates there is no notion of one dimension being larger than another since the metric has off-diagonal components. In order to see which dimensions become large it is necessary to go to a choice of coordinates where the generalized metric is diagonal. To simplify the notation we set $\rho=r^{D-4}$. Then the four eigenvalues of $\HH_{AB}$ are $\lambda_A$. They are given together with their limits by
\begin{equation}
\begin{array}{lccc}
& & r \rightarrow 0 & r \rightarrow \infty  \\ 
\lambda_1 = \phantom{-}\displaystyle\frac{h-f}{\rho}	& \quad\longrightarrow\quad	& 0	& -1 \\
\lambda_2 = -\displaystyle\frac{h-f}{\rho}	& \longrightarrow	& 0	& +1  \\
\lambda_3 = -\displaystyle\frac{h+f}{\rho}	& \longrightarrow	& -\infty			& -1 \\
\lambda_4 = \phantom{-}\displaystyle\frac{h+f}{\rho}	& \longrightarrow	& +\infty & +1
\end{array}
\end{equation}
where $f=\sqrt{h^2+\rho^2}$. The corresponding (normalized) eigenvectors can be used to construct the diagonalizing matrix which in turn is utilized to find the new basis where $\HH_{AB}$ is diagonal with entries $\lambda_A$. This new basis is given by 
\begin{equation}
{X'}^A =  \sqrt{\frac{h}{2f}}
\begin{pmatrix}
\sqrt{\frac{f+\rho}{h}} t - \sqrt{\frac{h}{f+\rho}} \tz \\
\sqrt{\frac{h}{f+\rho}} \ttt  + \sqrt{\frac{f+\rho}{h}} z \\
\sqrt{\frac{h}{f-\rho}} \ttt - \sqrt{\frac{f-\rho}{h}} z  \\
\sqrt{\frac{f-\rho}{h}} t + \sqrt{\frac{h}{f-\rho}} \tz
\end{pmatrix} \, .
\end{equation}
If we now take the limit where $r$ either goes to zero or infinity (and thus $f$ goes to $h$ or infinity), the diagonalized coordinate basis looks like (recall that the original basis was $X^A=(t,z,\ttt,\tz)$)
\begin{align}
r \rightarrow 0:\quad	{X'}^A &= \frac{1}{\sqrt{2}}
\begin{pmatrix}
t - \tz \\ \ttt + z \\  \ttt - z \\    t + \tz
\end{pmatrix}\, , &
r \rightarrow \infty: \quad {X'}^A &= 
\begin{pmatrix}
t \\  z  \\ \ttt \\ \tz
\end{pmatrix} \, .
\end{align}

This can be interpreted as follows. The DFT wave is asymptotically flat, thus for large $r$ there is no preferred set of coordinates, i.e. it does not matter which pair, $(t,z)$ or $(\ttt,\tz)$, we call ``dual" and which we take as being our usual spacetime. Different choices will just give T-duality related solutions. However, as one approaches the core of the solution and $r$ gets smaller, the space diagonalizes and takes the form of a sort of twisted light-cone. It is twisted in the sense that the light cone mixes the coordinates that at asymptotic infinity describe both the space and its dual. That is we have 
\begin{align}
u &= \frac{1}{\sqrt{2}}(t-\tz)\, , &
v &= \frac{1}{\sqrt{2}}(\ttt+z)\, , &
\tv &= \frac{1}{\sqrt{2}}(\ttt-z)\, , &
\tu &= \frac{1}{\sqrt{2}}(t + \tz) \, .
\end{align}

There is a clear hierarchy in the volumes between the two sets. One set, $\tv$ and $\tu$, comes with a large volume associated to those dimensions as can be seen from the eigenvalues $\lambda_3$ and $\lambda_4$ which diverge. On the other hand, the other set of coordinates, $u$ and $v$, is associated with a small volume as the eigenvalues $\lambda_1$ and $\lambda_2$ tend to zero. Thus one set is picked out and we must think of the core of the solution as being given by the space described by $\tv$ and $\tu$. Following these coordinates out of the core to asymptotic infinity these coordinates become just $\ttt$ and $\tz$. This fits in with the intuition that the dual description is suitable for describing the spacetime near the singular core. What may not have been apparent before DFT is that actually as one approaches the core these dual coordinates twist with the normal spacetime coordinates to give a twisted light-cone at the core of the fundamental string.

\subsection{Wave vs Monopole}
Now we return to the self-dual solution of EFT which is the main focus of this paper. To carry out the analysis, the solution is not treated as living in 4+56 dimensions but as a truely 60-dimensional solution. Thus the three constituents of the solution, the external metric $g_{\mu\nu}$, the extended internal metric $\MM_{MN}$ and the vector potential ${\AAA_\mu}^M$ are combined in the usual Kaluza-Klein fashion to form a 60-dimensional metric
\begin{equation}
\HH_{\hat{M}\hat{N}} = 
\begin{pmatrix}
g_{\mu\nu} + {\AAA_\mu}^M{\AAA_\nu}^N\MM_{MN} & {\AAA_\mu}^M\MM_{MN} \\
\MM_{MN}{\AAA_\nu}^N & \MM_{MN}
\end{pmatrix}
\end{equation}
where $g_{\mu\nu}$ is given in \eqref{eq:exmetric}, ${\AAA_\mu}^M$ in \eqref{eq:vecpot} and $\MM_{MN}$ is taken from \eqref{eq:genmetric} to be
\begin{equation}
\MM_{MN} = \diag[H^{3/2},H^{1/2}\delta_{27},H^{-1/2}\delta_{27},H^{-3/2}] \, .
\end{equation}
We then insert the EFT monopole/wave solution to find\footnote{The resulting object has dimension 60 whereas the two blocks have dimension 29 and 31 respectively as indicated by the indices $\hat{M}$, $A$ and $\bar{A}$.}
\begin{equation}
\HH_{\hat{M}\hat{N}} = 
\begin{pmatrix}
H^{1/2} \HH_{AB}^\mathrm{wave} & 0 \\
0 & H^{-1/2} \HH_{\bar{A}\bar{B}}^\mathrm{mono}
\end{pmatrix}
\label{eq:60metric}
\end{equation}
where to top left block is simply the metric of a wave with 27 transverse dimensions and the bottom right block is the metric of a monopole, also with 27 transverse dimensions
\begin{align}
\HH_{AB}^\mathrm{wave} &= 
\begin{pmatrix}
H-2 & H-1 & 0 \\
H-1 & H & 0 \\
0 & 0 & \delta_{27}
\end{pmatrix} , \\
\HH_{\bar{A}\bar{B}}^\mathrm{mono} &= 
\begin{pmatrix}
H(\delta_{ij} + H^{-2}A_iA_j) & H^{-1}A_i & 0 \\
H^{-1}A_j & H^{-1} & 0 \\
0 & 0 & \delta_{27}
\end{pmatrix} .
\end{align}
In \eqref{eq:60metric} it is interesting to see that there is a natural split into a block diagonal form simply by composing the fields \`a la Kaluza-Klein. These two blocks come with prefactors of $H^{\pm1/2}$ with opposite power, so the geometry will change distinctly between large and small $r$. 

Close to the core of the solution, where $r$ is small, $H$ becomes large and the wave geometry dominates. Far away for large $r$, $H$ will be close to one (and thus $A_i$ vanishes) and neither the monopole nor wave dominates. Thus one would imagine asymptotically either description is valid and the different choices related through a duality transformation. However what is curious is the dominance of the wave solution in the small $r$ region. It appears according to this analysis that all branes in string and M-theory when thought of as solutions in EFT are wave solutions in the core. It is hoped to explore these questions in a more rigorous fashion in future work.

\section*{Acknowledgement}
We thank Michael Duff and Malcolm Perry for comments on various aspects of this work. DSB is supported by the STFC consolidated grant ST/J000469/1 ``String Theory, Gauge Theory and Duality'' and FJR is supported by an STFC studentship.

\appendix

\section{Embedding the Type II Theories into EFT}
\label{app:embedding}

In this appendix we show how the Type II theories can be embedded into EFT. The difference between Type IIA and Type IIB arises from applying different solutions of the section condition to the EFT equations.

\subsection{The Type IIA Theory}
The ten-dimensional Type IIA theory is a simple reduction of eleven-dimensional supergravity on a circle. It is thus possible to embed it into EFT as well using the same solution to the section condition given in Section \ref{sec:embedding}. Instead of a $4+7$ coordinate split in the KK-decomposition we now have a $4+6$ split. The dictionary for embedding the Type IIA fields into EFT can be obtained from the results of Section \ref{sec:embedding} by simply splitting the internal seven-dimensional sector into $6+1$ by doing another KK-decomposition. 

Under the split $y^m = (y^\bm,y^\theta)$ with $\bm=1,\dots,6$ and $y^\theta=\theta$ the coordinate of the circle, the corresponding generalized coordinates read
\begin{equation}
Y^M = (y^\bm, y^\theta, y_{\bm\bn}, y_{\bm\theta} , 
		y_\bm, y_\btheta, y^{\bm\bn},y^{\bm\btheta}) \, .
\label{eq:gencoordIIA}
\end{equation}
Noting that the internal metric $g_{mn}$ of our solution is diagonal, no KK-vector will arise from this decomposition. We thus simply have
\begin{equation}
g_{mn} = \diag[ e^{-\phi/6}\bg_{\bm\bn}, e^{4\phi/3} ]
\end{equation}
where $e^{2\phi}$ is the string theory dilaton of the Type IIA theory and the precise numerical powers have been chosen to be in the Einstein frame. Inserting this ansatz into the generalized metric for embedding supergravity into EFT \eqref{eq:genmetricSUGRA} gives
\begin{align}
\MM_{MN}(\bg_{\bm\bn},\phi) = \bg^{1/2}\diag [&
	\bg_{\bm\bn}, e^{3\phi/2}, e^{\phi/2} \bg^{\bm\bn,\bk\bl}, e^{-\phi} \bg^{\bm\bn}, \notag\\
&	\bg^{-1} \bg^{\bm\bn}, e^{-3\phi/2}\bg^{-1}, e^{-\phi/2}\bg^{-1}\bg_{\bm\bn,\bk\bl},
		e^{\phi}\bg^{-1}\bg_{\bm\bn}]
\label{eq:genmetricIIA}
\end{align}
where $\bg$, $\bg_{\bm\bn,\bk\bl}$ and $\bg^{\bm\bn,\bk\bl}$ are defined in terms of $\bg_{\bm\bn}$ as before (barred quantities are six-dimensional). As in the eleven-dimensional case, we are not considering any internal components of the RR or NSNS gauge potentials. The only non-zero components are the one-form parts which are in the EFT vector potential ${\AAA_\mu}^M$ as before. The vector is also split under the above decomposition resulting in a component for each of the directions given in \eqref{eq:gencoordIIA}
\begin{equation}
\left\{ {\AAA_\mu}^\bm, {\AAA_\mu}^\theta, \AAA_{\mu\ \bm\bn}, \AAA_{\mu\ \bm\theta},
\AAA_{\mu\ \bm}, \AAA_{\mu\ \theta}, {\AAA_\mu}^{\bm\bn}, {\AAA_\mu}^{\bm\theta} \right\} \, .
\label{eq:vecpotIIA}
\end{equation}
All these parts of ${\AAA_\mu}^M$ encode a component of a field from the Type IIA theory except $\AAA_{\mu\ \bm}$ which relates to the dual graviton. The first one, ${\AAA_\mu}^\bm$ is just the KK-vector of the original $4+6$ decomposition. The RR-fields $C_1$, $C_3$, $C_5$ and $C_7$ are encoded in ${\AAA_\mu}^\theta$, $\AAA_{\mu\ \bm\bn}$, ${\AAA_\mu}^{\bm\bn}$ and $\AAA_{\mu\ \theta}$ respectively where the latter two have to be dualized on the internal six-dimensional space. The remaining two, $\AAA_{\mu\ \bm\theta}$ and ${\AAA_\mu}^{\bm\theta}$ contain the NSNS-fields $B_2$ and $B_6$, where again the second one has to be dualized. It is nice to see how the self-duality of the EFT vector contains all the known dualities between the form fields in the Type IIA theory.

\subsection{The Type IIB Theory}
Unlike the Type IIA theory, the Type IIB theory does not follow from the solution to the section condition that gives eleven-dimensional supergravity. There is another, inequivalent solution \cite{Hohm:2013uia} which is related to a different decomposition of the fundamental representation of $E_7$. The relevant maximal subgroup is $GL(6)\times SL(2)$ and we have
\begin{equation}
\mathbf{56} \rightarrow (6,1) + (6,2) + (20,1) + (6,2) + (6,1)
\end{equation}  
which translates to the following splitting of the extended internal coordinates
\begin{equation}
Y^M = (y^\bm, y_{\bm\ a}, y_{\bm\bn\bk}, y^{\bm\ a}, y_\bm)
\label{eq:gencoordIIB}
\end{equation}
where again $\bm=1,\dots,6$ and $a=1,2$ is an $SL(2)$ index. The middle component is totally antisymmetric in all three indices. Note that the six-dimensional index is \emph{not} the same as in the $6+1$ Type IIA decomposition above. Here we rather have a $5+2$ split where the $y_{ab}$ (a single component) is reinterpreted as the sixth component of $y^\bm$. Loosely speaking this comes from the fact that Type IIB on a circle is related to M-theory on a torus. This is made precise at the end of this section.

From \cite{Hohm:2013uia}, the generalized metric (again without any contribution from the internal components of the form fields) for this case is given by
\begin{equation}
\MM_{MN}(\bg_{\bm\bn},\gamma_{ab}) = \bg^{1/2}\diag [ \bg_{\bm\bn}, \bg^{\bm\bn}\gamma^{ab}, \bg^{-1}\bg_{\bm\bk\bp,\bn\bl\bq},
	\bg^{-1}\bg_{\bm\bn}\gamma_{ab}, \bg^{-1}\bg^{\bm\bn}]
\label{eq:genmetricIIB}
\end{equation}
where $\bg_{mkp,nlq}=\bg_{m[n|}\bg_{k|l|}\bg_{p|q]}$ (in analogy to $g_{mn,kl}$ above) and $\gamma_{ab}$ is the metric on the torus
\begin{equation}
\gamma_{ab} = \frac{1}{\mathrm{Im\, } \tau}
\begin{pmatrix}
|\tau|^2 & \mathrm{Re\, } \tau \\ \mathrm{Re\, } \tau & 1
\end{pmatrix},
\qquad
\tau = C_{0} + i e^{-\phi}
\label{eq:gamma}
\end{equation}
with the complex torus parameter $\tau$ (the ``axio-dilaton'') given in terms of the RR-scalar $C_0$ and the string theory dilaton $e^{2\phi}$. We will come back to this setup at the end of this section.
 
The EFT vector is also decomposed and has a component for each direction in \eqref{eq:gencoordIIB}
\begin{equation}
\left\{ {\AAA_\mu}^\bm, \AAA_{\mu\ \bm \ a}, \AAA_{\mu\ \bm\bn\bk},
	{\AAA_\mu}^{\bm\ a}, \AAA_{\mu\ \bm} \right\} \, .
\label{eq:vecpotIIB}
\end{equation}
As before, these parts each encode a component of a field from the Type IIB theory except $\AAA_{\mu\ \bm}$ which relates to the dual graviton. As always, ${\AAA_\mu}^\bm$ is the KK-vector of the original $4+6$ decomposition. The components $\AAA_{\mu\ \bm \ a}$ and ${\AAA_\mu}^{\bm\ a}$ contain the $SL(2)$ doublets $B_2/C_2$ and $B_6/C_6$ where the latter one needs to be dualized on the internal space. Here $B$ denotes a NSNS-field and $C$ the dual RR-field. The component $\AAA_{\mu\ \bm\bn\bk}$ corresponds to the self-dual four-form $C_4$. Again it can be seen that the self-duality of the EFT vector gives the duality relations between the form fields in the Type IIB theory.

Let us conclude by checking how the Type IIB theory on a circle is related to the eleven-dimensional theory on a torus. Since both theories have the external four-dimensional spacetime in common, we will only look at the internal sector. The seven dimensions are split into $5+2$ such that the coordinates are $y^m=(y^\dm,y^a)$ where $\dm=1,\dots,5$ and $a=1,2$. Starting from \eqref{eq:gencoordM}, the generalized coordinates then decompose as
\begin{equation}
Y^M = (y^\dm, y^a, y_{\dm\dn}, y_{\dm a}, y_{ab}, 
	y_\dm, y_a, y^{\dm\dn}, y^{\dm a}, y^{ab})\, .
\end{equation}
By noting that $y_{ab}$ has only a single component (by antisymmetry), $y_{12}$, these coordinates can be repackaged into $y^\bm=(y^\dm,y_{12})$ and similar for the dual coordinates to make contact with \eqref{eq:gencoordIIB}. We thus have
\begin{equation}
\begin{array}{rrl}
(6,1): 		& 	y^\bm &= (y^\dm,y_{12}) \\
(6,2): 		&	y_{\bm a} &= (y_{\dm a}, y^a)\\
(20,1):		&	y_{\bm\bn\bk} &= (y_{\dm\dn},y^{\dm\dn})\\
(6,2):		&	y^{\bm a} &= (y^{\dm a}, y_a) \\
(6,1):		&	y_\bm &= (y_\dm,y^{12})
\end{array}
\end{equation}
justifying the presence of the six-dimensional index $\bm$ above. Now turn to the seven-dimensional metric $g_{mn}$. Again omitting a KK-vector for cross-terms, the ansatz for the decomposition is (a dot denotes a five-dimensional quantity)
\begin{equation}
g_{mn} = \diag[ \dg_{\dm\dn}, e^\Delta \gamma_{ab} ]
\end{equation}
with $\gamma_{ab}$ as given in \eqref{eq:gamma}. There the torus metric is conformal and has unit determinant. For completeness, we include a volume factor for the torus in the discussion here, such that the determinant of the $2\times 2$ sector $g_{ab}$ is $\det|e^\Delta\gamma_{ab}| =e^{2\Delta}$. This ansatz can be inserted into the generalized metric for embedding supergravity into EFT \eqref{eq:genmetricSUGRA} to give
\begin{align}
\MM_{MN} = \dg^{1/2} \diag[& e^\Delta\dg_{\dm\dn},e^{2\Delta}\gamma_{ab}, 
		e^\Delta\dg^{\dm\dn,\dpp\dq},
		\dg^{\dm\dn}\gamma^{ab}, e^{-\Delta}\gamma^{ab,cd}, \notag\\
&		\dg^{-1}e^{-\Delta}\dg^{\dm\dn}, \dg^{-1}e^{-2\Delta}\gamma^{ab},
		\dg^{-1}e^{-\Delta}\dg_{\dm\dn,\dpp\dq}, \\
&		\dg^{-1}\dg_{\dm\dn}\gamma_{ab},
		\dg^{-1}e^\Delta\gamma_{ab,cd}] \, . \notag
\end{align}
It is easy to check that the object $\gamma_{ab,cd}=\gamma_{a[c}\gamma_{d]b}$ has only one component $\gamma_{12,12}$ which evaluates to $1$ (and similarly for the inverse $\gamma^{ab,cd}$). With this in mind, the components of the generalized metric can be repackaged in terms of a six-dimensional metric $g_{\bm\bn}=e^{\Delta/2}\diag[\dg_{\dm\dn},e^{-2\Delta}]$ and determinant $\bg=e^\Delta\dg$ just like the coordinates. The five parts of the generalized metric thus read
\begin{equation}
\begin{array}{rrll}
(6,1): 		& 	\bg_{\bm\bn} &= e^{\Delta/2}&\diag[\dg_{\dm\dn},e^{-2\Delta}] \\
(6,2): 		&	\bg^{\bm\bn}\gamma^{ab} &= e^{-\Delta/2}&\diag[\dg^{\dm\dn}\gamma^{ab}, 		
												e^{2\Delta}\gamma_{ab}] \\
(20,1):		&	\bg^{-1}\bg_{\bm\bk\bp,\bn\bl\bq} &= e^{\Delta/2}&\diag[\dg^{\dm\dn,\dpp\dq}, 		
												e^{-2\Delta}\dg^{-1}\dg_{\dm\dn,\dpp\dq}] \\
(6,2):		&	\bg^{-1}\bg_{\bm\bn}\gamma_{ab} &= e^{-\Delta/2} 
												&\diag[\dg^{-1}\dg_{\dm\dn}\gamma_{ab}, 		
												e^{-2\Delta}\dg^{-1}\gamma^{ab}] \\
(6,1):		&	\bg^{-1}\bg^{\bm\bn} &= e^{-3\Delta/2}&\diag[\dg^{-1}\dg_{\dm\dn},
												\dg^{-1}e^{2\Delta}]
\end{array}
\end{equation}
which is in agreement with \eqref{eq:gencoordIIB}. These identifications here are not obvious, but can be checked by an explicit calculation of individual components.

\section{Glossary of Solutions}
\label{app:glossary}

The purpose of this appendix is not only to collect all the fundamental, solitonic and Dirichlet solutions of ten- and eleven-dimensional supergravities as they can be found in any standard text book (for us Ortin's book \cite{Ortin04} was an invaluable source), but also to present them with their fields rearranged according to a Kaluza-Klein coordinate split. It is the decomposed fields that are extracted from the EFT solution in the main text. It also highlights some interesting similarities between these solutions, such as that they \emph{all} have the same four-dimensional external spacetime under the decomposition.

The coordinates $\hx^\hmu=(x^\mu,x^m)$ are either split into $11\rightarrow 4+7$ or $10\rightarrow 4+6$ and the corresponding KK-decomposition takes the form 
\begin{equation}
\hg_{\hmu\hnu} = 
			\begin{pmatrix}
				g_{\mu\nu} + {A_\mu}^m{A_\nu}^ng_{mn} & {A_\mu}^mg_{mn} \\
				g_{mn}{A_\nu}^n & g_{mn}
			\end{pmatrix}
\end{equation}
where hatted quantities are ten- or eleven-dimensional and the internal sector is six- or seven-dimensional. The off-diagonal or cross-term ${A_\mu}^m$ is called the KK-vector and will mostly be zero except for the wave and the monopole. The four-dimensional external metric $g_{\mu\nu}$ has to be rescaled by the determinant of the internal metric $g_{mn}$ to remain in the Einstein frame. This is crucial for comparing solutions and takes the form
\begin{equation}
g_{\mu\nu} \rightarrow |\det g_{mn}|^{1/2} g_{\mu\nu} \, .
\end{equation}
The power of the determinant in the rescaling depends on the number of external dimensions and is $1/2$ in our case.

The eleven-dimensional supergravity solutions are specified in terms of the metric $\hg_{\hmu\hnu}$ and the three-form and the six-form potentials $C_3$ and $C_6$ which are duals of each other. In the NSNS-sector, the fields of the ten-dimensional Type II solutions are the metric $\hg_{\hmu\hnu}$, the string theory dilaton\footnote{The constant part of the dilaton is denoted by $e^{2\phi_0}$ where $\phi_0$ is a constant which can be set to zero if convenient.} $e^{2\phi}$ and the two-form and six-form Kalb-Ramond potentials $B_2$ and $B_6$ which again are duals. In the RR-sector we have the $C_p$ potentials with $p=1,\dots,7$ in this paper. The odd ones belong to the Type IIA theory and the even ones to the Type IIB theory.

From an EFT point of view, the external metric is simply the rescaled $g_{\mu\nu}$. The form fields and the KK-vector ${A_\mu}^m$ constitute the components of the EFT vector ${\AAA_\mu}^M$. The generalized metric $\MM_{MN}$ is constructed from the internal metric $g_{mn}$ according to \eqref{eq:genmetricSUGRA}. The dilaton $\phi$ in Type IIA or the axio-dilaton $\tau$ in Type IIB also enter the generalized metric as in \eqref{eq:genmetricIIA} and \eqref{eq:genmetricIIB} respectively.

Each solution is presented with its full field content in terms of the harmonic function $H$ which has a functional dependence on the transverse directions of each solution. Then we perform the explained KK-decomposition by picking time and three of the transverse directions to be in the four-dimensional external sector and the world volume directions together with the remaining transverse ones to be in the six- or seven-dimensional internal sector. As part of the decomposition the solution is \emph{smeared} over those transverse directions in the internal sector so that it is only localized in the three transverse directions in the external sector, i.e. $H=1+h/|r|$ with $r^2=\delta_{ij}w^iw^j$ and the $w$'s denote these three directions.

A final note on the notation: $t$ is the time coordinate, $z$ is the ``special'' direction of the wave and the monopole, $\vec{x}_{(p)}$ denotes the $p$ world volume directions of a p-brane and $\vec{y}_{(D-1-p)}$ the remaining $D-1-p$ transverse directions, the first three of which are usually taken to be in the external sector as explained above, i.e. $w^i=y^i$ for $i,=1,2,3$.

\subsection{Wave, Membrane, Fivebrane and Monopole in $D=11$}
\label{app:SUGRA}

\subsubsection*{The Wave - WM}
\begin{equation}
\begin{aligned}
\dd s^2 &= -H^{-1}\dd t^2 + H\left[\dd z - (H^{-1}-1)\dd t\right]^2 
				+ \dd\vec{y}_{(9)}^{\, 2}   \\
		&= (H-2)\dd t^2 + 2(H-1)\dd t \dd z + H\dd z^2 + \dd\vec{y}_{(9)}^{\, 2} \\
H &= 1 + \frac{h}{|\vec{y}_{(9)}|^7}  
\end{aligned}
\label{eq:WM}
\end{equation}
KK-decomposition: $x^\mu = (t, \vec{y}_{(3)})$ and $x^m = (z, \vec{y}_{(6)})$
\begin{equation}
\begin{aligned}
g_{mn} &= \diag[H,\delta_6] \, , & \det g_{mn}&=H  \\
g_{\mu\nu} &= \diag[-H^{-1/2},H^{1/2}\delta_{ij}] \, , &  {A_t}^z &= -(H^{-1}-1) 
\end{aligned}
\end{equation}

\subsubsection*{The Membrane - M2}
\begin{equation}
\begin{aligned}
\dd s^2 &= H^{-2/3}[-\dd t^2 + \dd\vec{x}_{(2)}^{\, 2}] + H^{1/3}\dd\vec{y}_{(8)}^{\, 2}   \\
C_{tx^1x^2} &= -(H^{-1}-1), \qquad C_{iy^4y^5y^6y^7y^8} = A_i\\
H &= 1 + \frac{h}{|\vec{y}_{(8)}|^6} 
\end{aligned}
\label{eq:M2}
\end{equation}
KK-decomposition: $x^\mu = (t, \vec{y}_{(3)})$ and $x^m = (\vec{x}_{(2)}, \vec{y}_{(5)})$
\begin{equation}
\begin{aligned}
g_{mn} &= H^{1/3}\diag[H^{-1}\delta_2,\delta_5] \, , & \det g_{mn}&=H^{1/3}  \\
g_{\mu\nu} &= \diag[-H^{-1/2},H^{1/2}\delta_{ij}] 
\end{aligned}
\end{equation}

\subsubsection*{The Fivebrane - M5}
\begin{equation}
\begin{aligned}
\dd s^2 &= H^{-1/3}[-\dd t^2 + \dd\vec{x}_{(5)}^{\, 2}] + H^{2/3}\dd\vec{y}_{(5)}^{\, 2}   \\
C_{tx^1x^2x^3x^4x^5} &= -(H^{-1}-1),  \qquad C_{iy^4y^5} = A_i  \\
H &= 1 + \frac{h}{|\vec{y}_{(5)}^{\, 2}|^3}  \, .
\end{aligned}
\label{eq:M5}
\end{equation}
KK-decomposition: $x^\mu = (t, \vec{y}_{(3)})$ and $x^m = (\vec{x}_{(5)}, \vec{y}_{(2)})$
\begin{equation}
\begin{aligned}
g_{mn} &= H^{2/3}\diag[H^{-1}\delta_5,\delta_2] \, , & \det g_{mn}&=H^{-1/3}  \\
g_{\mu\nu} &= \diag[-H^{-1/2},H^{1/2}\delta_{ij}] 
\end{aligned}
\end{equation}

\subsubsection*{The Monopole - KK7}
\begin{equation}
\begin{aligned}
\dd s^2 &= -\dd t^2 + \dd\vec{x}_{(6)}^{\, 2} 
				+ H^{-1}\left[\dd z + A_i\dd y^i\right]^2 + H\dd\vec{y}_{(3)}^{\, 2}  \\
		&= -\dd t^2 + \dd\vec{x}_{(6)}^{\, 2} + H^{-1}\dd z^2 + 2H^{-1}A_i\dd y^i\dd z
				+ H\left(\delta_{ij}+H^{-2}A_iA_j\right)\dd y^i\dd y^j \\
H &= 1 + \frac{h}{|\vec{y}_{(3)}|} \, , \qquad
\partial_{[i}A_{j]} = \frac{1}{2}{\epsilon_{ij}}^k\partial_k H \, .
\end{aligned}
\label{eq:KK7}
\end{equation}
KK-decomposition: $x^\mu = (t, \vec{y}_{(3)})$ and $x^m = (z, \vec{x}_{(6)})$
\begin{equation}
\begin{aligned}
g_{mn} &= \diag[H^{-1},\delta_6] \, , & \det g_{mn}&=H^{-1}  \\
g_{\mu\nu} &= \diag[-H^{-1/2},H^{1/2}\delta_{ij}] \, , &  {A_i}^z &= A_i 
\end{aligned}
\end{equation}

\subsubsection*{The M2/M5 Bound State}
\begin{equation}
\begin{aligned}
\dd s^2 &= H^{-2/3}\Xi^{1/3}[\dd t^2 + \dd\vec{x}_{(2)}^{\, 2}] 
			+ H^{1/3}\Xi^{1/3}\dd\vec{y}_{(5)}^{\, 2}
			+ H^{1/3}\Xi^{-2/3}\dd\vec{z}_{(3)}^{\, 2}   \\
C_{tx^1x^2} &= -(H^{-1}-1)\sin\xi,  \qquad C_{iy^4y^5z^1x^2z^3} = A_i\sin\xi  \\
C_{iy^4y^5} &= A_i\cos\xi,  \qquad C_{tx^1z^2z^1x^2z^3} = -(H^{-1}-1)\cos\xi \\ 
C_{z^1z^2z^3} &= -(H-1)\Xi^{-1}\sin\xi\cos\xi \\
H &= 1 + \frac{h}{|\vec{y}_{(3)}|},  \qquad\Xi = \sin^2\xi + H \cos^2\xi \, .
\end{aligned}
\label{eq:M2M5}
\end{equation}
KK-decomposition: $x^\mu = (t, \vec{y}_{(3)})$ and $x^m = (\vec{x}_{(2)}, \vec{y}_{(2)}, \vec{z}_{(3)})$
\begin{equation}
\begin{aligned}
g_{mn} &= H^{1/3}\Xi^{1/3}\diag[H^{-1}\delta_2,\delta_2,\Xi^{-1}\delta_3] \, , & 
\det g_{mn}&=H^{1/3}\Xi^{-2/3}  \\
g_{\mu\nu} &= \diag[-H^{-1/2},H^{1/2}\delta_{ij}] 
\end{aligned}
\end{equation}

\subsection{Wave, String, Fivebrane and Monopole in $D=10$}
\label{app:Strings}

\subsubsection*{The Wave - WA/B}
\begin{equation}
\begin{aligned}
\dd s^2 &= -H^{-1}\dd t^2 + H\left[\dd z - (H^{-1}-1)\dd t\right]^2 
				+ \dd\vec{y}_{(8)}^{\, 2}   \\
H &= 1 + \frac{h}{|\vec{y}_{(8)}|^6} \, ,  \qquad e^{2\phi}=e^{2\phi_0}
\end{aligned}
\label{eq:WA}
\end{equation}
KK-decomposition: $x^\mu = (t, \vec{y}_{(3)})$ and $x^\bm = (z, \vec{y}_{(5)})$
\begin{equation}
\begin{aligned}
\bg_{\bm\bn} &= \diag[H,\delta_5] \, , & \det \bg_{\bm\bn}& = H \\
g_{\mu\nu} &= \diag[-H^{-1/2},H^{1/2}\delta_{ij}] \, , &  {A_t}^z &= -(H^{-1}-1) 
\end{aligned}
\end{equation}

\subsubsection*{The Fundamental String - F1}
\begin{equation}
\begin{aligned}
\dd s^2 &= H^{-3/4}[-\dd t^2 + \dd x^2] + H^{1/4}\dd\vec{y}_{(8)}^{\, 2}   \\
B_{tx} &= -(H^{-1}-1), \qquad B_{iy^4y^5y^6y^7y^8} = A_i\\
H &= 1 + \frac{h}{|\vec{y}_{(8)}|^6} , \qquad e^{2\phi} = H^{-1}e^{2\phi_0}
\end{aligned}
\label{eq:F1}
\end{equation}
KK-decomposition: $x^\mu = (t, \vec{y}_{(3)})$ and $x^\bm = (x, \vec{y}_{(5)})$
\begin{equation}
\begin{aligned}
\bg_{\bm\bn} &= H^{1/4}\diag[H^{-1},\delta_5] \, , & \det \bg_{\bm\bn}&=H^{1/2}  \\
g_{\mu\nu} &= \diag[-H^{-1/2},H^{1/2}\delta_{ij}] 
\end{aligned}
\end{equation}

\subsubsection*{The Solitonic Fivebrane - NS5}
\begin{equation}
\begin{aligned}
\dd s^2 &= H^{-1/4}[-\dd t^2 + \dd\vec{x}_{(5)}^{\, 2}] + H^{3/4}\dd\vec{y}_{(4)}^{\, 2}   \\
B_{tx^1x^2x^3x^4x^5} &= -(H^{-1}-1), \qquad B_{iy^4} = A_i\\
H &= 1 + \frac{h}{|\vec{y}_{(4)}|^2} , \qquad e^{2\phi} = He^{2\phi_0}
\end{aligned}
\label{eq:NS5}
\end{equation}
KK-decomposition: $x^\mu = (t, \vec{y}_{(3)})$ and $x^\bm = (\vec{x}_{(5)}, y^4)$
\begin{equation}
\begin{aligned}
\bg_{\bm\bn} &= H^{3/4}\diag[H^{-1}\delta_5, 1] \, , & \det \bg_{\bm\bn}&=H^{-1/2}  \\
g_{\mu\nu} &= \diag[-H^{-1/2},H^{1/2}\delta_{ij}] 
\end{aligned}
\end{equation}

\subsubsection*{The Monopole - KK6A/B}
\begin{equation}
\begin{aligned}
\dd s^2 &= -\dd t^2 + \dd\vec{x}_{(5)}^{\, 2} 
				+ H^{-1}\left[\dd z + A_i\dd y^i\right]^2 + H\dd\vec{y}_{(3)}^{\, 2}  \\
H &= 1 + \frac{h}{|\vec{y}_{(3)}|} \, , \qquad e^{2\phi} = e^{2\phi_0}
\end{aligned}
\label{eq:KK6A}
\end{equation}
KK-decomposition: $x^\mu = (t, \vec{y}_{(3)})$ and $x^\bm = (z, \vec{x}_{(5)})$
\begin{equation}
\begin{aligned}
\bg_{\bm\bn} &= \diag[H^{-1},\delta_5] \, , & \det \bg_{\bm\bn}&=H^{-1}  \\
g_{\mu\nu} &= \diag[-H^{-1/2},H^{1/2}\delta_{ij}] \, , &  {A_i}^z &= A_i 
\end{aligned}
\end{equation}

\subsection{D-Branes in $D=10$}
\label{app:Dbranes}

\subsubsection*{The Dp-Brane for $p=0,\dots,6$}
\begin{equation}
\begin{aligned}
\dd s^2 &= H^{\frac{p-7}{8}}[-\dd t^2 + \dd\vec{x}_{(p)}^{\, 2}] 
			+ H^{\frac{p+1}{8}}\dd\vec{y}_{(9-p)}^{\, 2}   \\
C_{tx^1\dots x^p} &= -(H^{-1}-1), \qquad C_{iy^4\dots y^{9-p}} = A_i\\
H &= 1 + \frac{h}{|\vec{y}_{(9-p)}|^{7-p}} , \qquad e^{2\phi} = H^{\frac{3-p}{2}} e^{2\phi_0}
\end{aligned}
\label{eq:Dbrane}
\end{equation}
KK-decomposition: $x^\mu = (t, \vec{y}_{(3)})$ and $x^\bm = (\vec{x}_{(p)}, \vec{y}_{(6-p)})$
\begin{equation}
\begin{aligned}
\bg_{\bm\bn} &= H^{\frac{p+1}{8}}\diag[H^{-1}\delta_p,\delta_{6-p}] \, , & 
	\det \bg_{\bm\bn}&=H^{\frac{3-p}{4}}  \\
g_{\mu\nu} &= \diag[-H^{-1/2},H^{1/2}\delta_{ij}] 
\end{aligned}
\end{equation}
Note: In Type IIB the D1-brane forms an S-duality doublet with the F1-string. This means they are identical solutions up to an $SL(2)$ transformation and their dilatons are inverses of each other. The same applies for the D5-brane and the NS5-brane.

\addcontentsline{toc}{section}{References}
\bibliographystyle{JHEP} 
\bibliography{mybib}

\end{document}